\pdfoutput=1
\documentclass[twocolumn]{aastex62}

\usepackage{comment}

\usepackage{graphicx}
\usepackage{txfonts}
\usepackage{hyperref}

\shorttitle{Peculiarities of NOAA AR 12673}
\shortauthors{Getling}

\begin{document}

\title{Peculiarities of the Dynamics of Solar NOAA Active Region 12673}

\correspondingauthor{A. V. Getling}
\email{A.Getling@mail.ru}

\author{A. V. Getling}
\affil{Skobeltsyn Institute of Nuclear Physics, Lomonosov Moscow State
              University, Moscow, 119991 Russia}

 \begin{abstract}
       The dynamics of active region (AR) 12673 is qualitatively studied using observational data obtained with the Helioseismic and Magnetic Imager of the \emph{Solar Dynamics Observatory} on August~31--September~8, 2017. This AR was remarkable for its complex structure and extraordinary flare productivity. The sunspot group in this AR consisted of (1) an old,  well-developed and highly stable, coherent sunspot, which had also been observed two solar rotations earlier, and (2) a rapidly developing cluster of umbral and penumbral fragments. Cluster (2) formed two elongated, arc-shaped chains of spot elements, skirting around the major sunspot (1), with two chains of magnetic elements spatially coinciding with the arcs. AR components (1) and (2) were in relative motion, cluster (2) overtaking spot (1) in westward motion, and their relative velocity agrees in order of magnitude with the velocity jump over the near-surface shear layer, or leptocline. The pattern of motion of the features about the main spot bears amazing resemblance to the pattern of a fluid flow about a roundish body. This suggests that spot (1) was dynamically coupled with the surface layers, while cluster (2) developed in deeper layers of the convection zone. The magnetic-flux emergence in cluster (2) appeared to be associated with fluid motions similar to roll convection. The mutual approach of components (1) and (2) gave rise to light bridges in the umbrae of sunspots with the magnetic field having the same sign on both sides of the bridge.
\end{abstract}


\section{Introduction}

\begin{figure*}
\centering
\includegraphics[width=0.3\textwidth]
{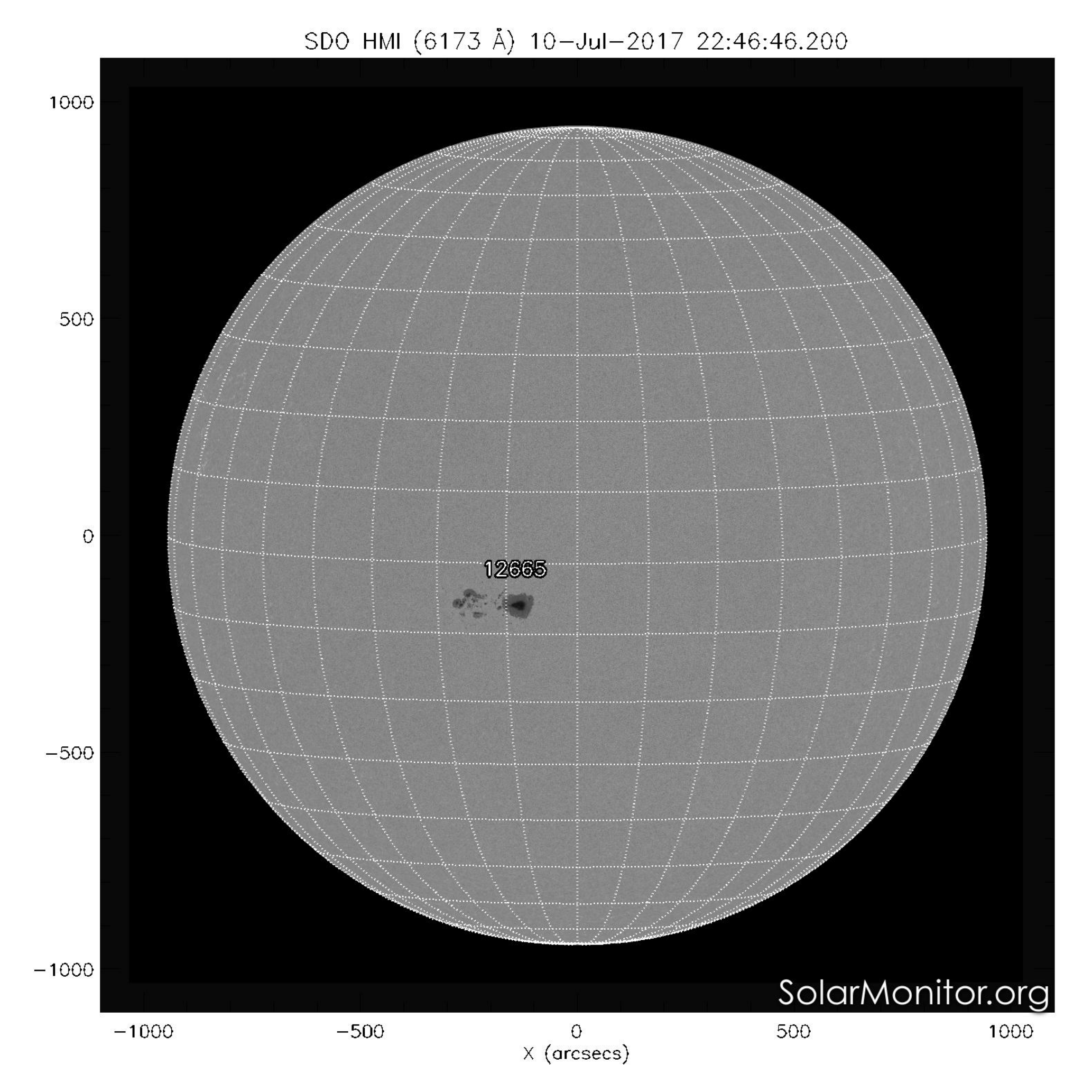}
\includegraphics[width=0.3\textwidth]
{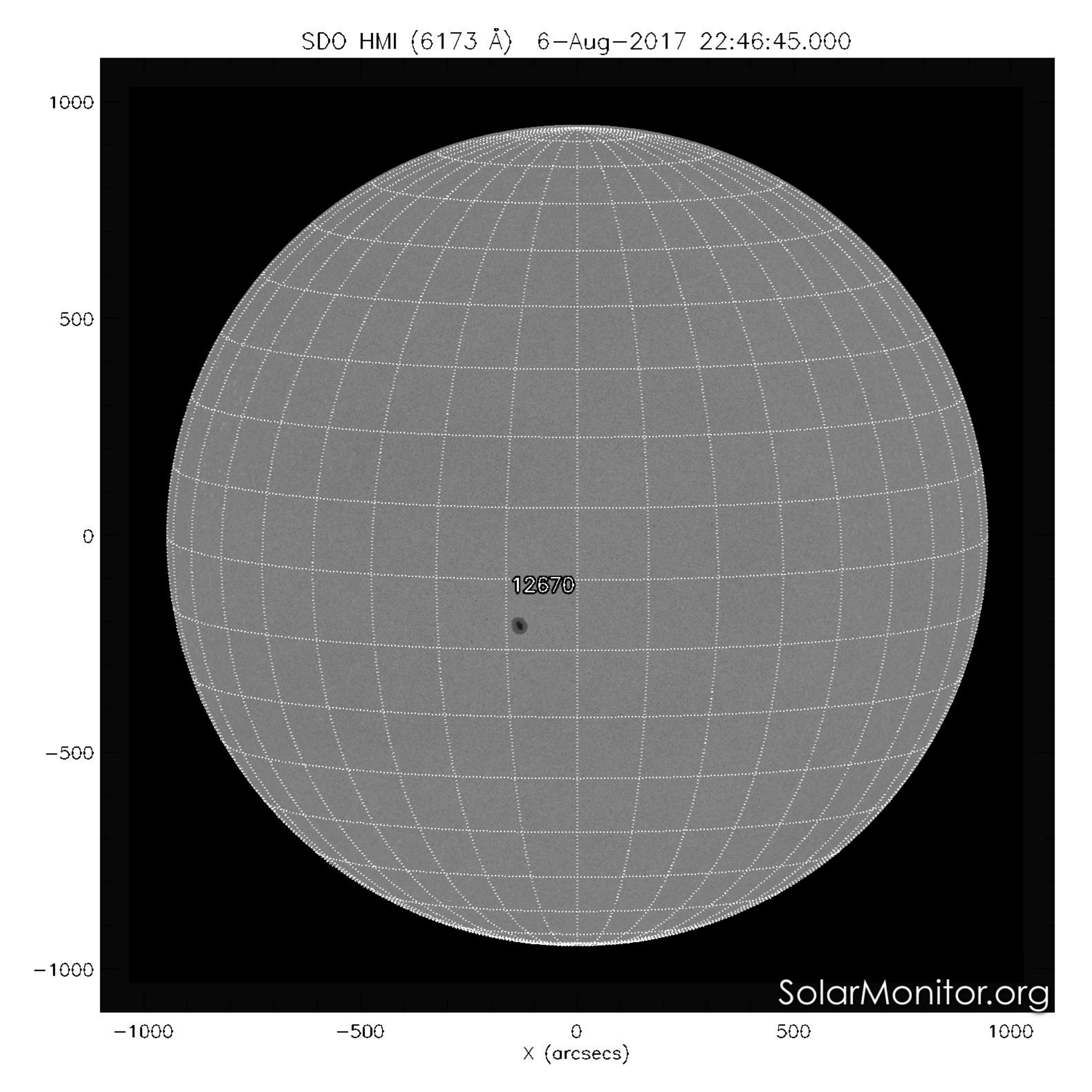}\\
\tiny{\hspace{0.4cm}(a)\hspace{5.3cm}(b)}\\
\includegraphics[width=0.3\textwidth]
{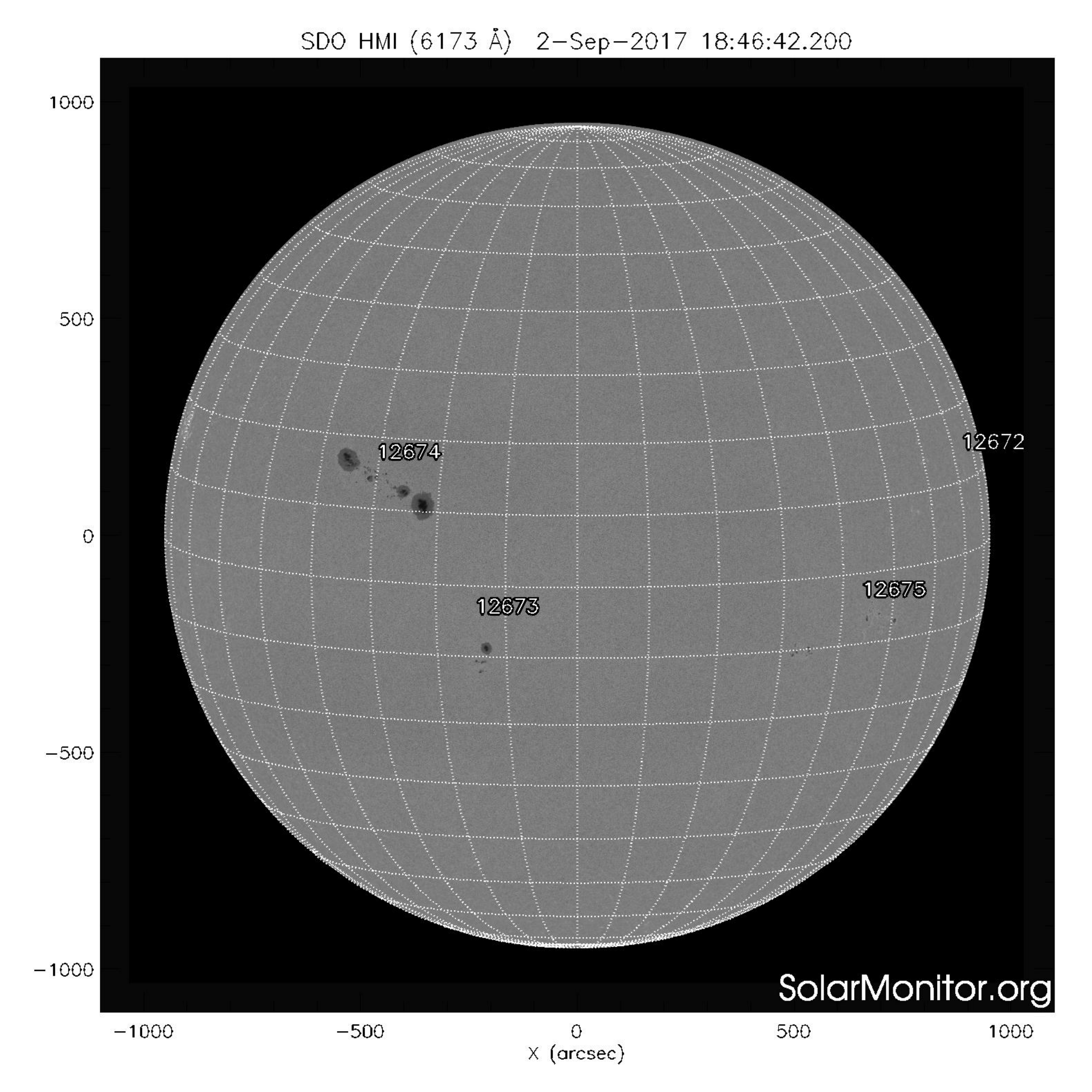}
\includegraphics[width=0.3\textwidth]
{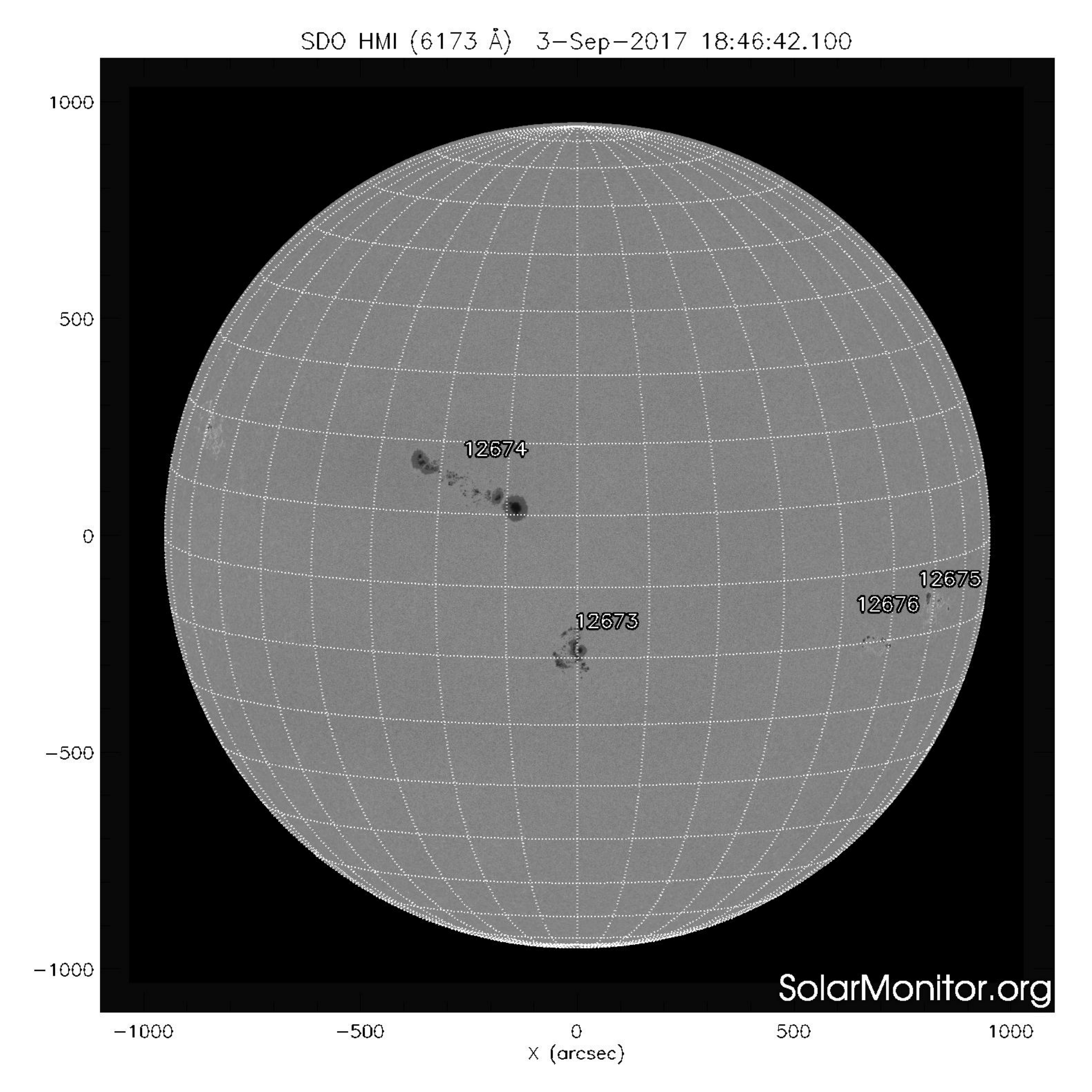}
\includegraphics[width=0.3\textwidth]
{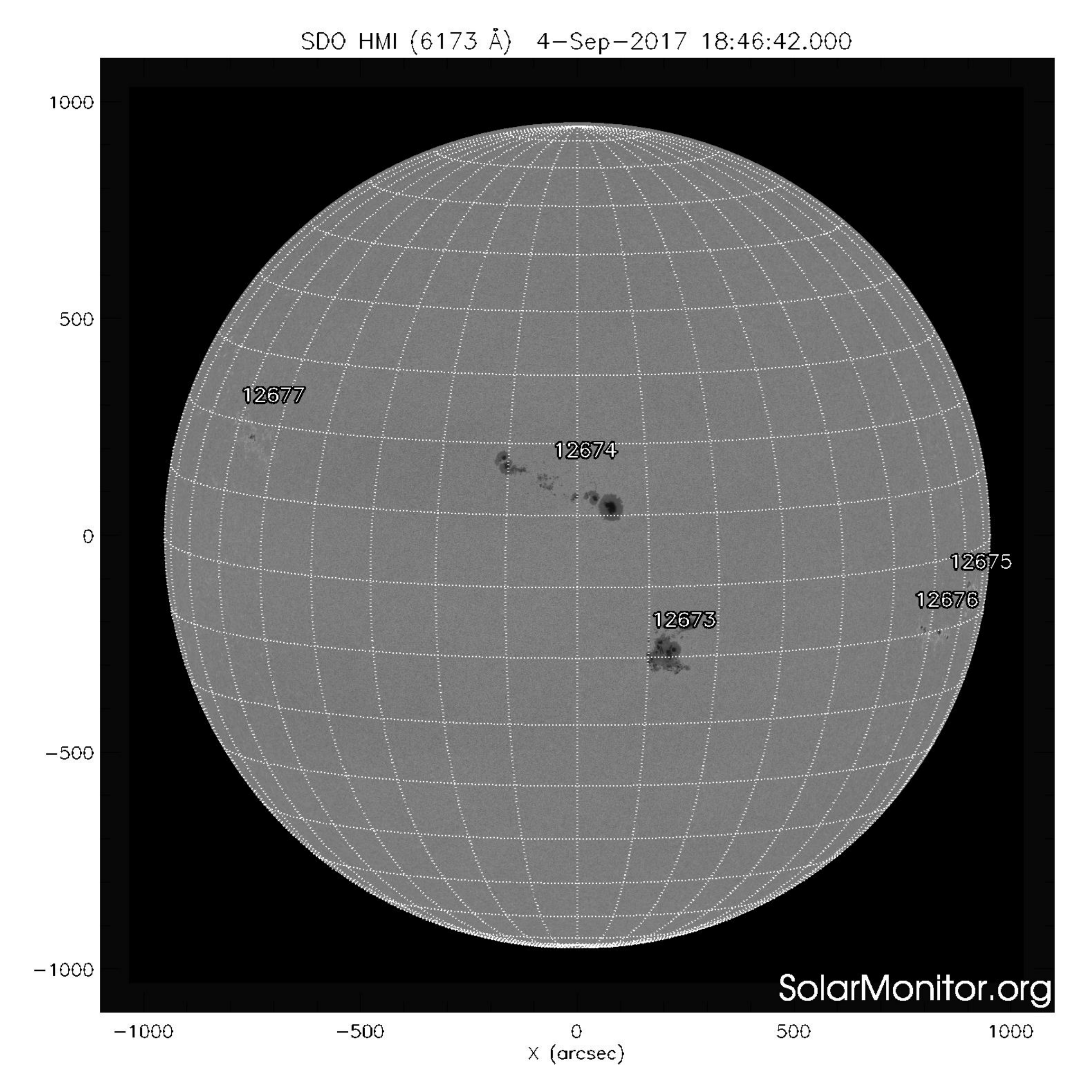}\\
\tiny{\hspace{0.4cm}(c)\hspace{5.3cm}(d)\hspace{5.3cm}(e)}\\
\includegraphics[width=0.3\textwidth]
{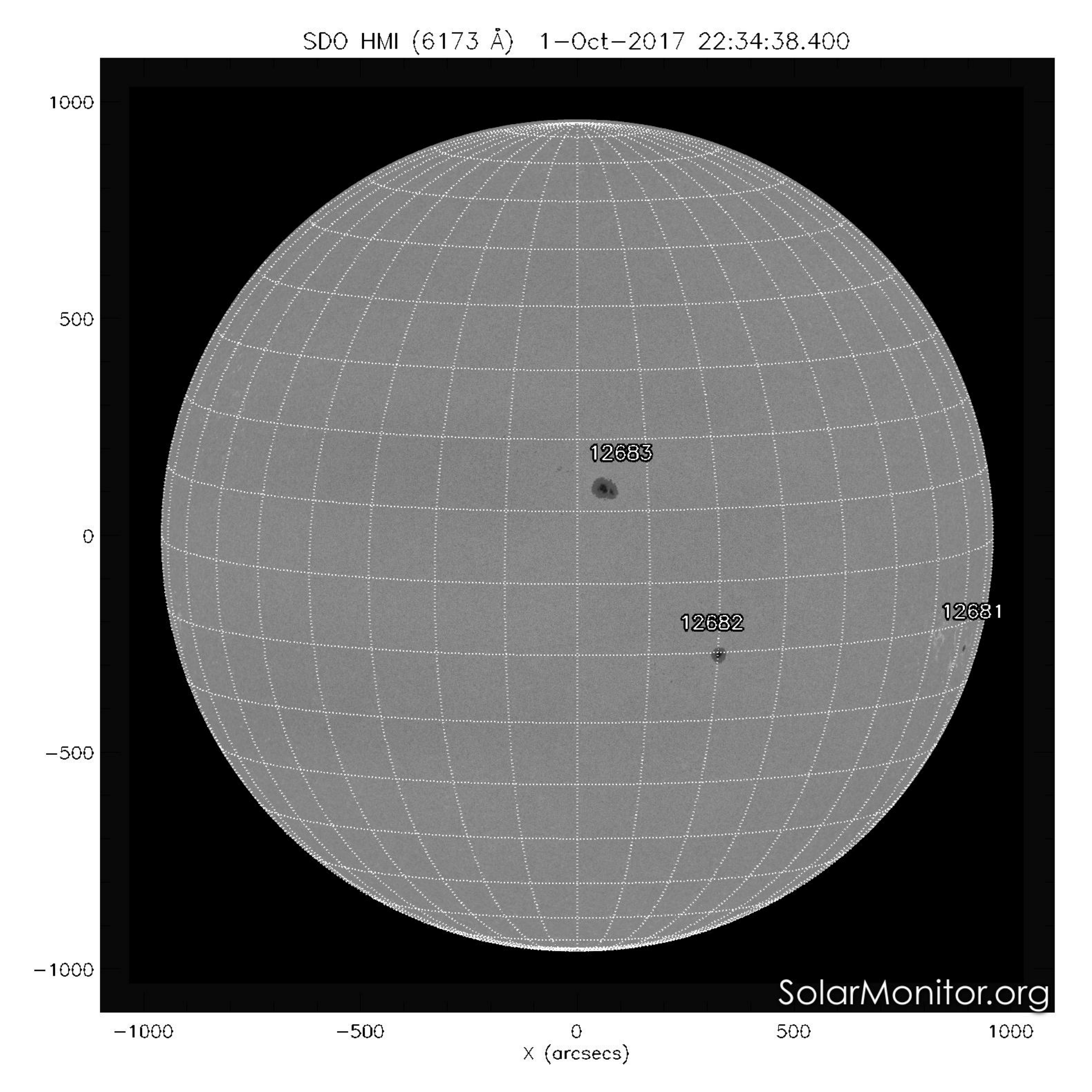}
\includegraphics[width=0.3\textwidth]
{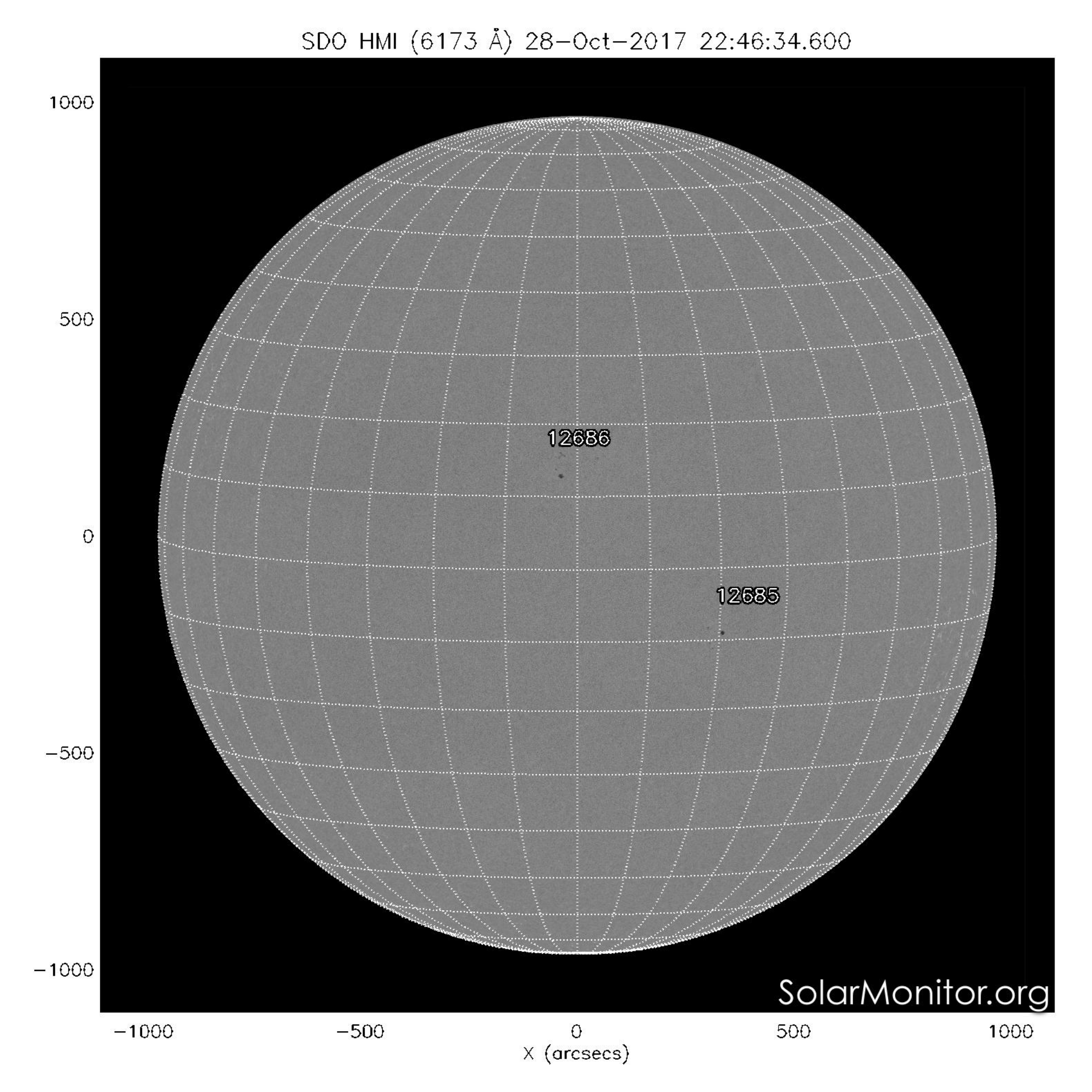}\\
\tiny{\hspace{0.4cm}(f)\hspace{5.3cm}(g)}\\
\caption{Prehistory, history, and posthistory of AR 12673 (full-disk photospheric images from \url{https://solarmonitor.org}): (a) sunspot group in AR 12665; (b) sunspot in AR 12670 at a time of one synodic rotation after (a); (c--e) sunspot pattern including a group in AR 12673 at three times taken 1 day apart, starting from a time of one synodic rotation after (b); (f) sunspot pattern including a spot in AR 12682 at a time of one synodic rotation after (e); (g) sunspot pattern including a spot in AR 12685 at a time of one synodic rotation after (f).}
\label{history}
\end{figure*}

Solar NOAA active region (AR) 12673, which was observed in September 2017, was remarkable for its complex structure and extraordinary flare productivity---the highest in Solar Cycle 24 \citep{Yang_etal_2017,Attie_etal_2018, Hou_etal_2018}. It produced 4 flares above X1 class and 8 flares above M3 class; moreover, its X-9.3 flare (on 2017 September 6, 11:53 UT) was the most intense one since 2005 \citep[see, e.g.,][]{Sun_Norton_2017}. The behavior of the flow and magnetic field in this AR deserves special attention. In particular, the magnetic-field and sunspot dynamics suggests that different parts of this complex aggregate might be dynamically coupled with different layers of the solar convection zone and atmosphere.

The sunspot group in AR~12673 consisted of (1) a well-developed, highly stable, coherent sunspot with a magnetic field exceeding 2~kG, which had previously existed, and (2) a cluster of umbral and penumbral fragments, which emerged on 2017 September 2 and fused later into new larger spots. We will see that component (2) was not definitely opposite to spot (1) in its polarity, so that the whole magnetic configuration of AR 12673 could not be classified as a bipolar one. Sunspot (1), as its position on the solar disk suggests, was already present two solar rotations before cluster (2) started developing (Figure~\ref{history}a), being the leading spot of a regular bipolar group attributed to AR~12665. One rotation later, this sunspot appeared to be unipolar, belonging to AR~12670 (Figure~\ref{history}b).

The evolution of AR~12673 (Figures~\ref{history}c--\ref{history}e) as the main subject of this study will be discussed below. Now, we only note that a unipolar sunspot was also present one and two rotations after the period of development of AR 12673 not far from its location, being attributed to ARs~12682 and 12685 (Figures~\ref{history}f and \ref{history}g, respectively).

This study follows the avenue of investigation of the AR and sunspot-group formation mechanisms as outlined by \cite{Getling_Buchnev2019} and based on parallel analyses of the magnetic and velocity fields simultaneously recorded in an AR. Currently available observational facilities provide a wealth of data admitting multifaceted description of the MHD processes in question. This should be highly promising in terms of {understanding} the physical mechanisms involved.

\begin{figure*}
\centering
\includegraphics[width=0.275\textwidth,bb=0 2 456 435,clip]
{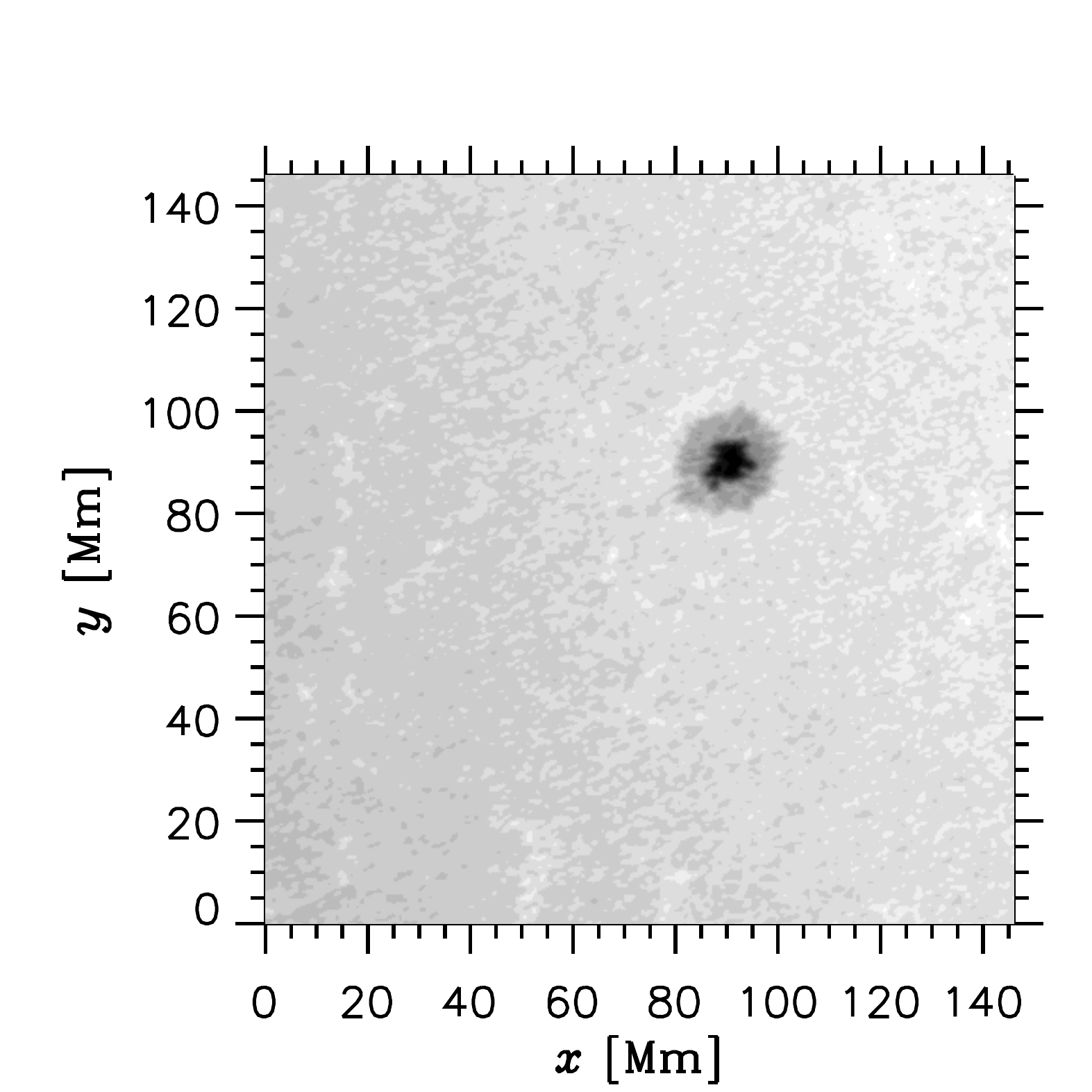}
\includegraphics[width=0.3142\textwidth,bb=0 -8 589 435,clip]
{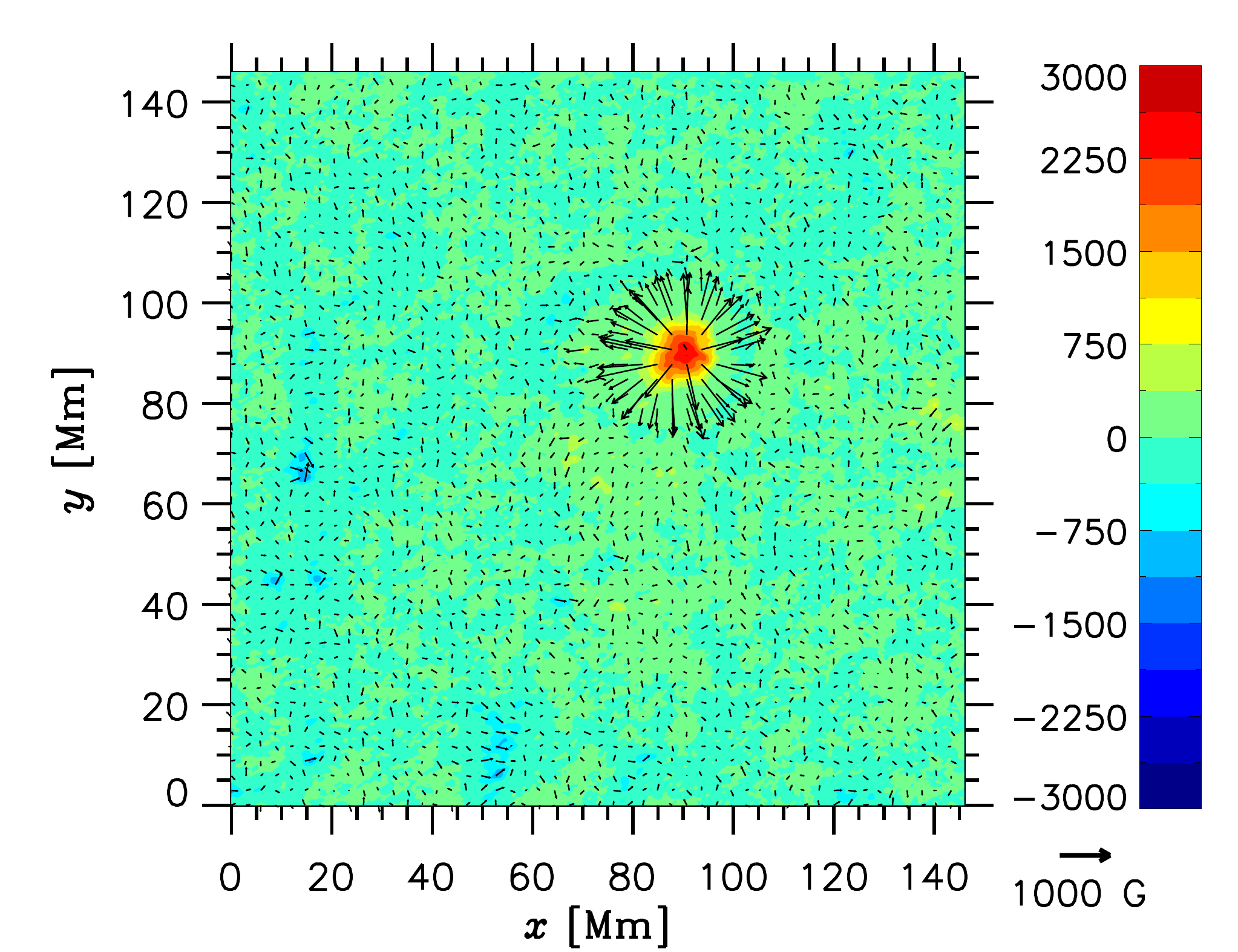}\\[-2pt]
\tiny{2017 August 31, 00:00 TAI}\\[-12pt]
\includegraphics[width=0.275\textwidth,bb=0 2 456 435,clip] {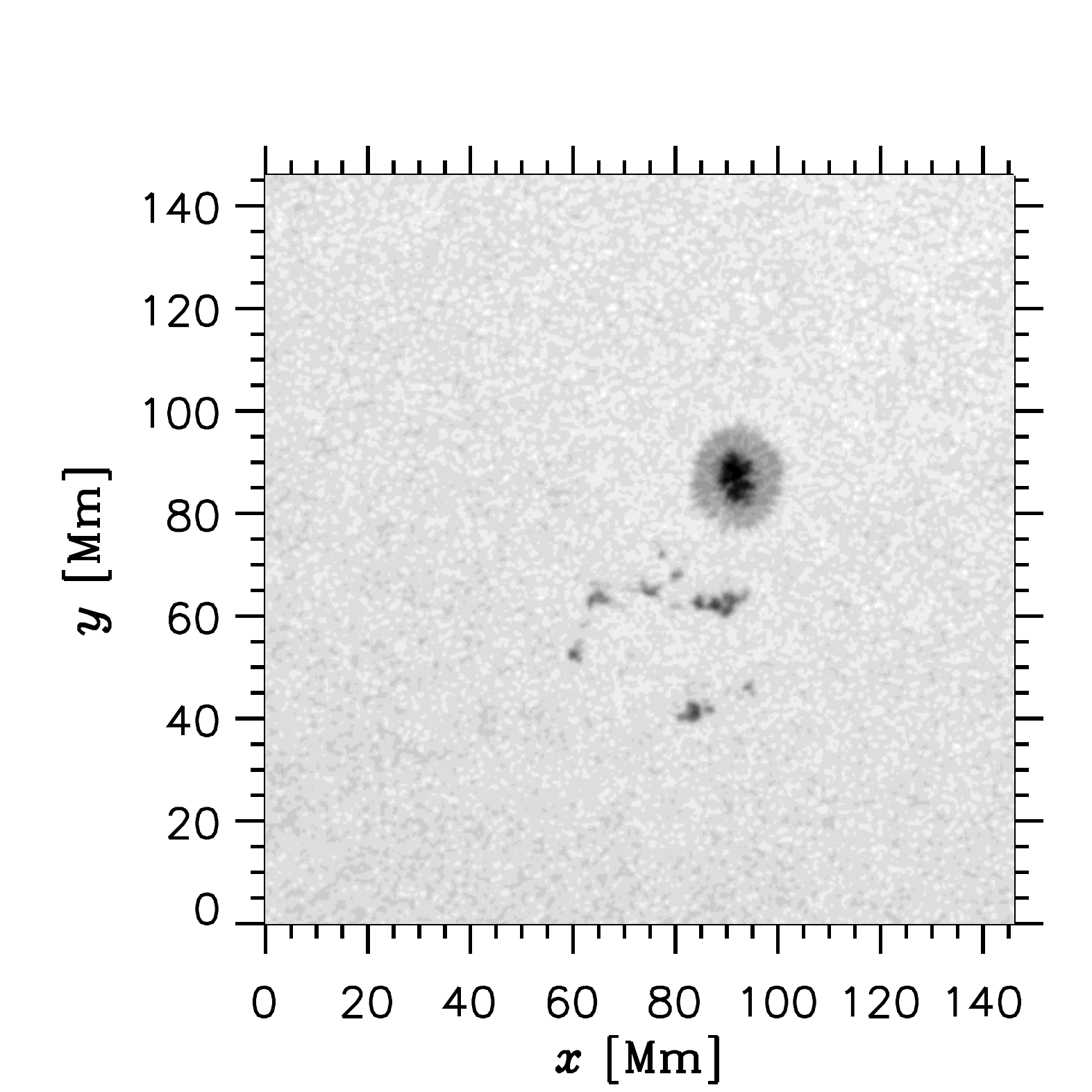}
\includegraphics[width=0.3142\textwidth,bb=0 -8 589 435,clip] {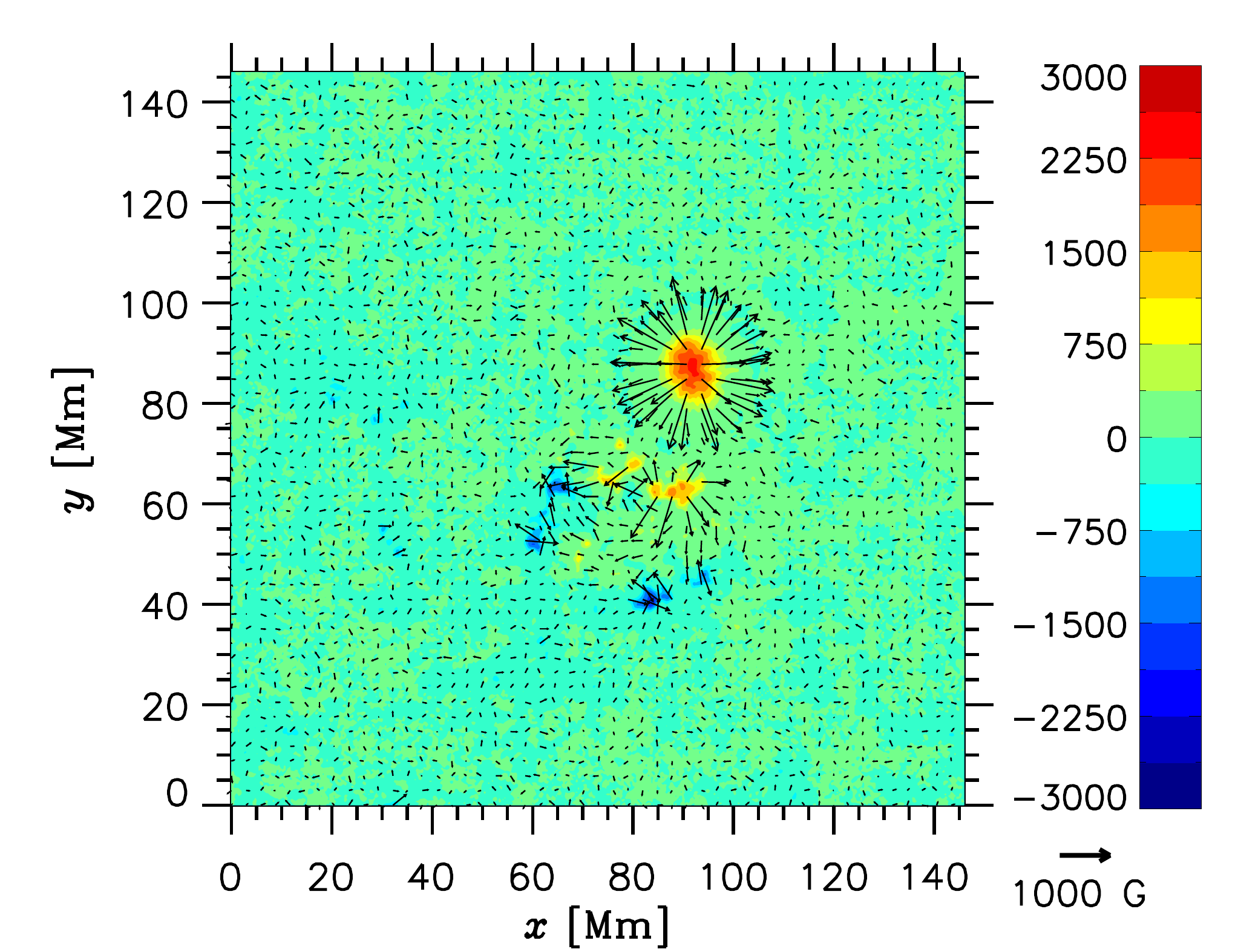}\\[-2pt]
\tiny{2017 September 03, 00:00 TAI}\\[-12pt]
\includegraphics[width=0.275\textwidth,bb=0 2 456 435,clip]
{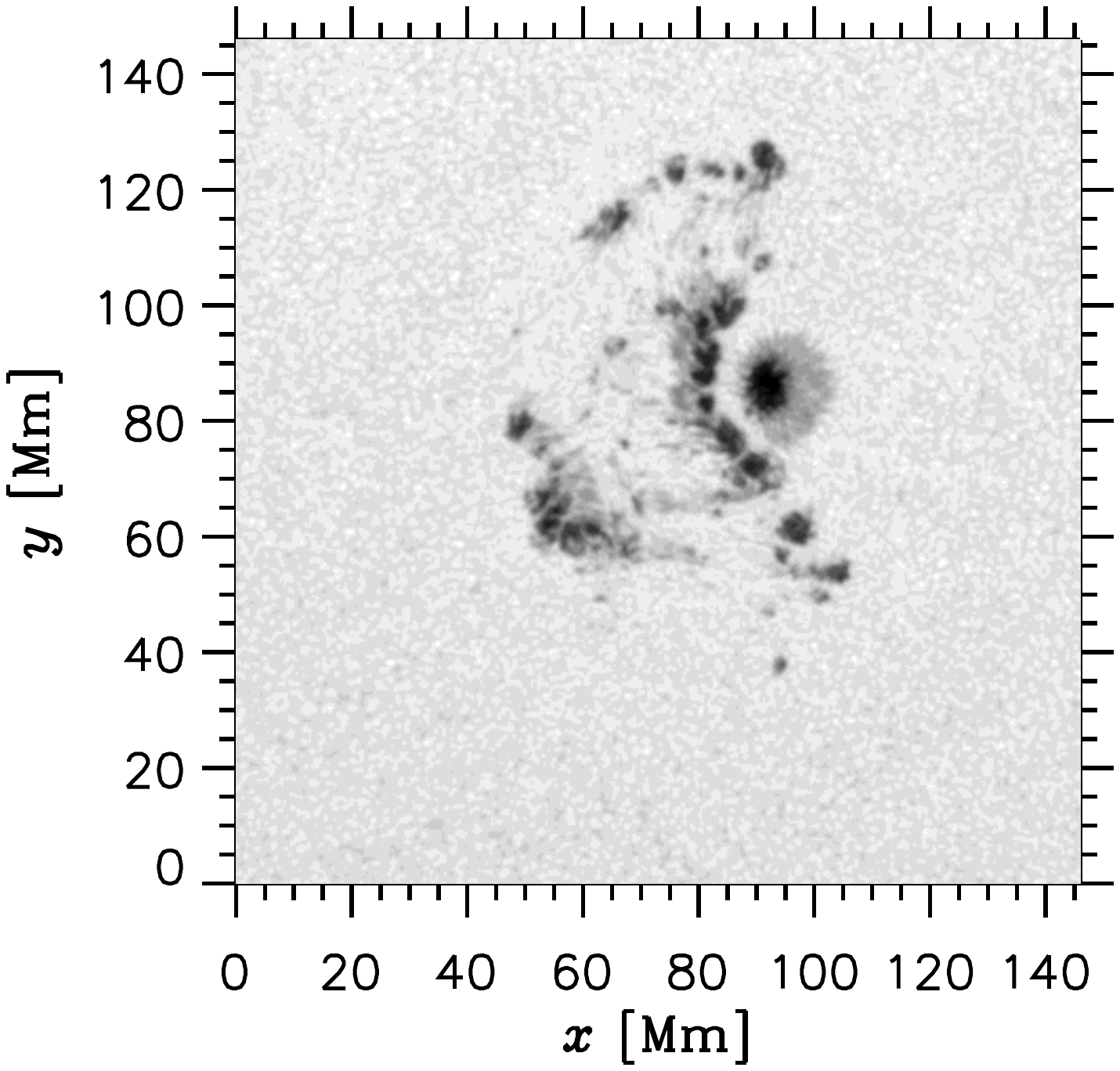}
\includegraphics[width=0.3142\textwidth,bb=0 -8 589 435,clip]
{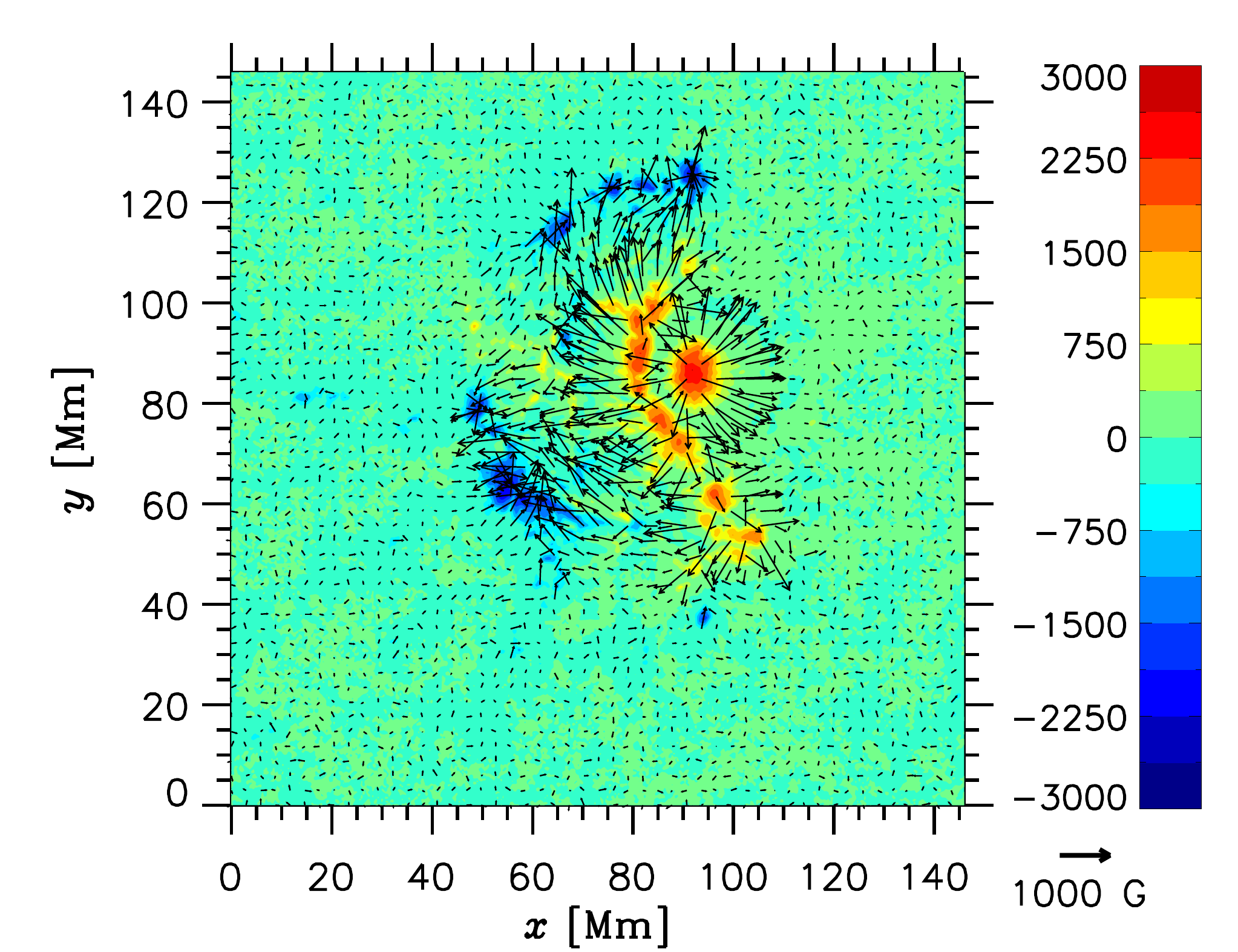}\\[-2pt]
\tiny{2017 September 03, 18:00 TAI}\\[-12pt]
\includegraphics[width=0.275\textwidth,bb=0 2 456 435,clip] {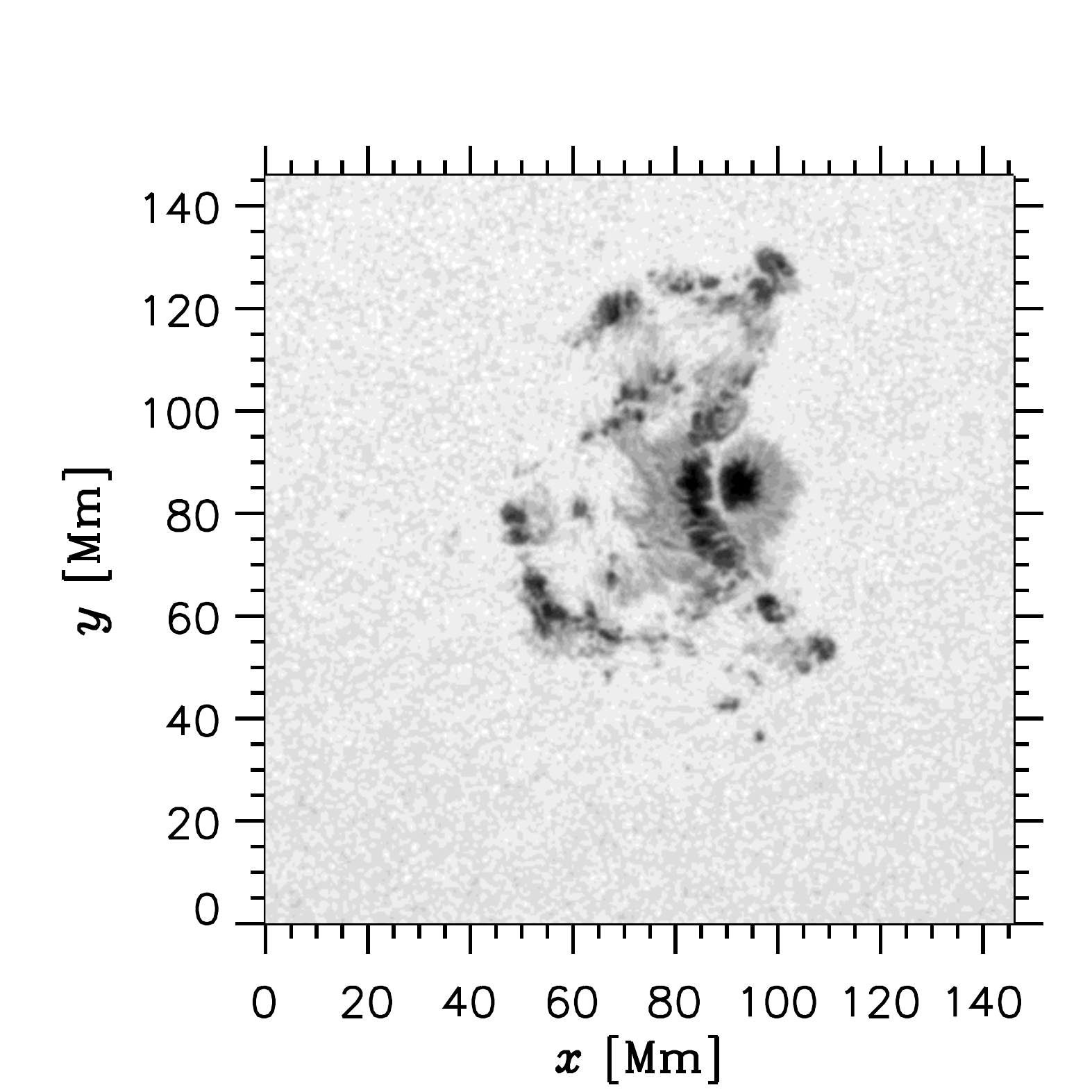}
\includegraphics[width=0.3142\textwidth,bb=0 -8 589 435,clip] {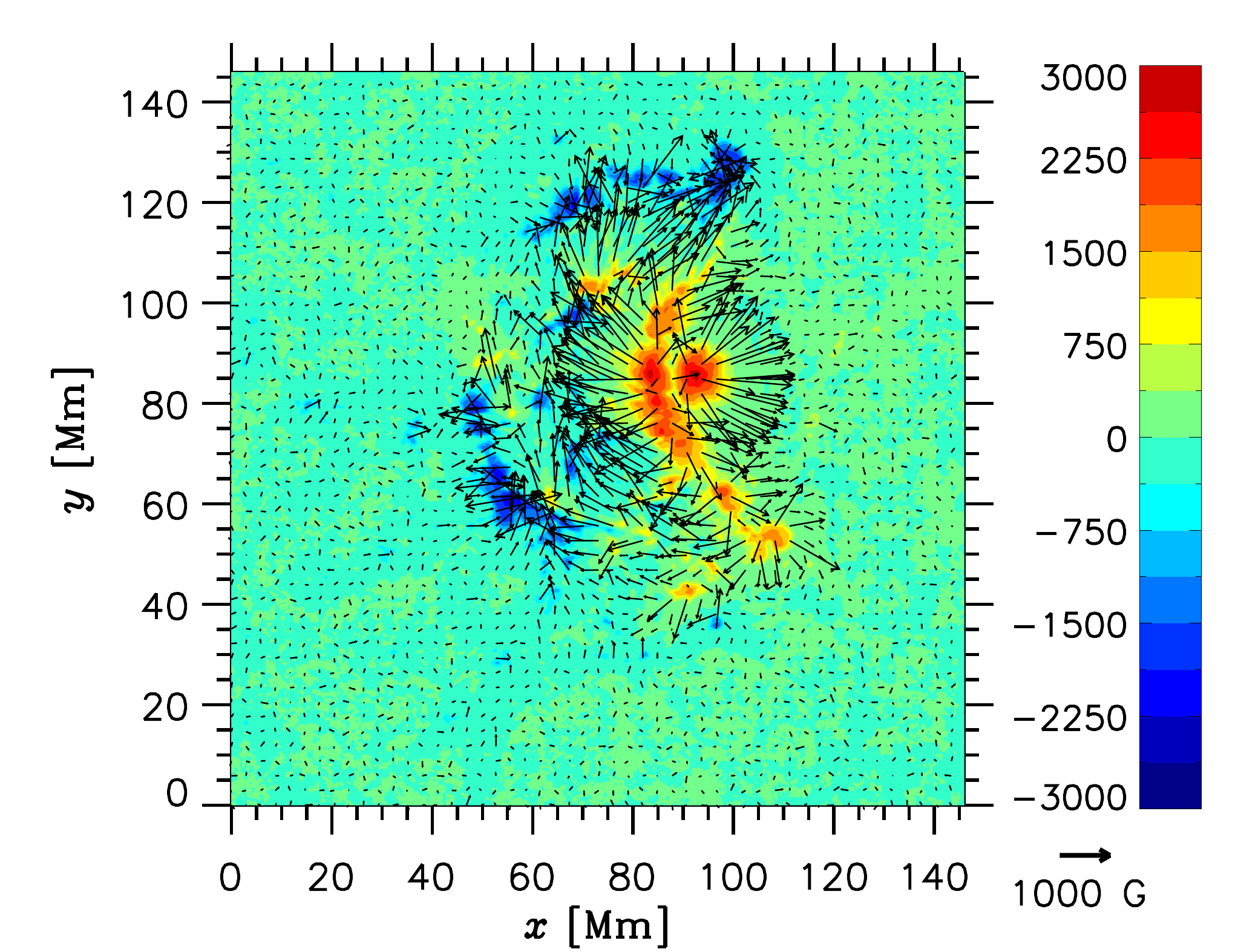}\\[-2pt]
\tiny{2017 September 04, 00:00 TAI}\\[-12pt]
\includegraphics[width=0.275\textwidth,bb=0 2 456 435,clip]
{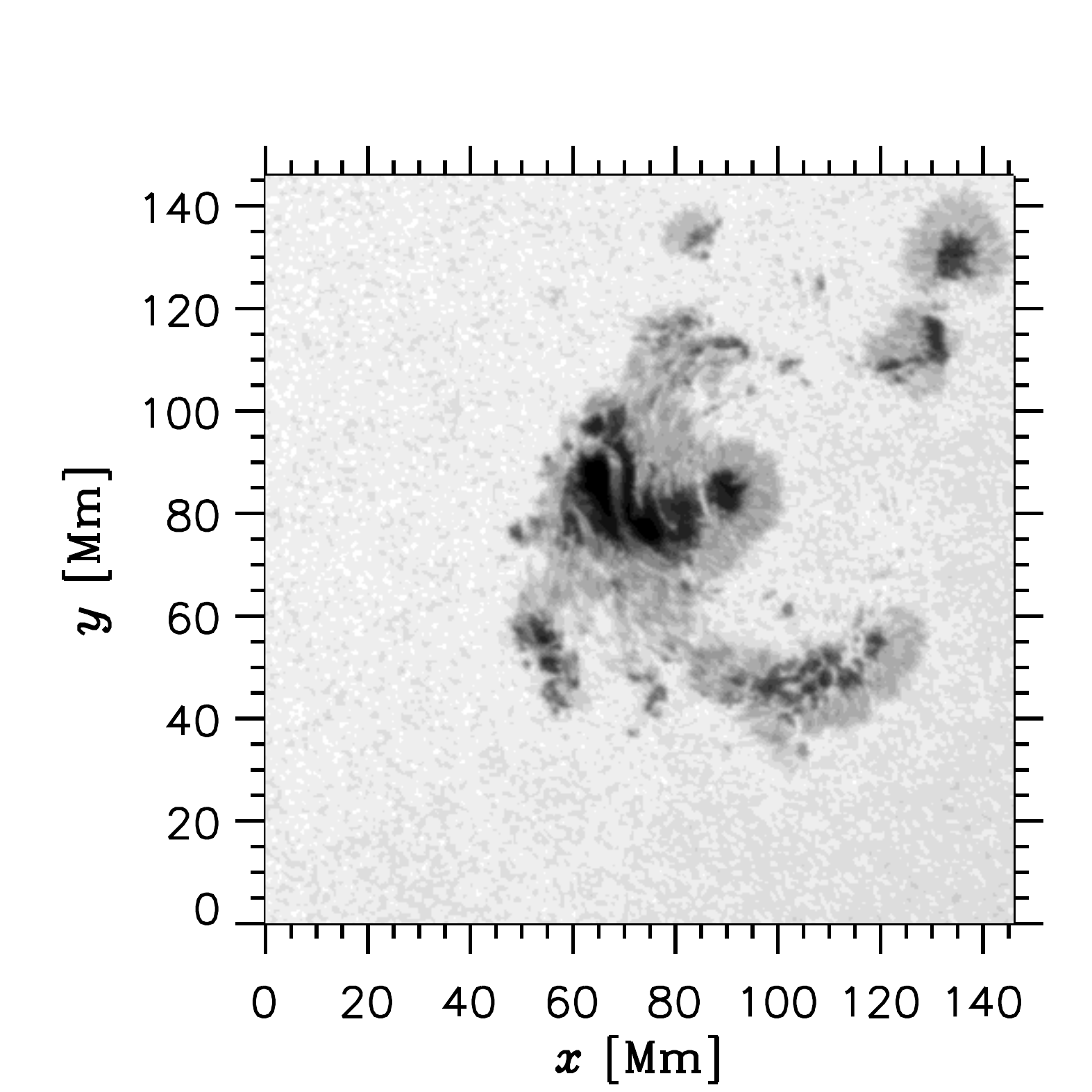}
\includegraphics[width=0.3142\textwidth,bb=0 -8 589 435,clip]
{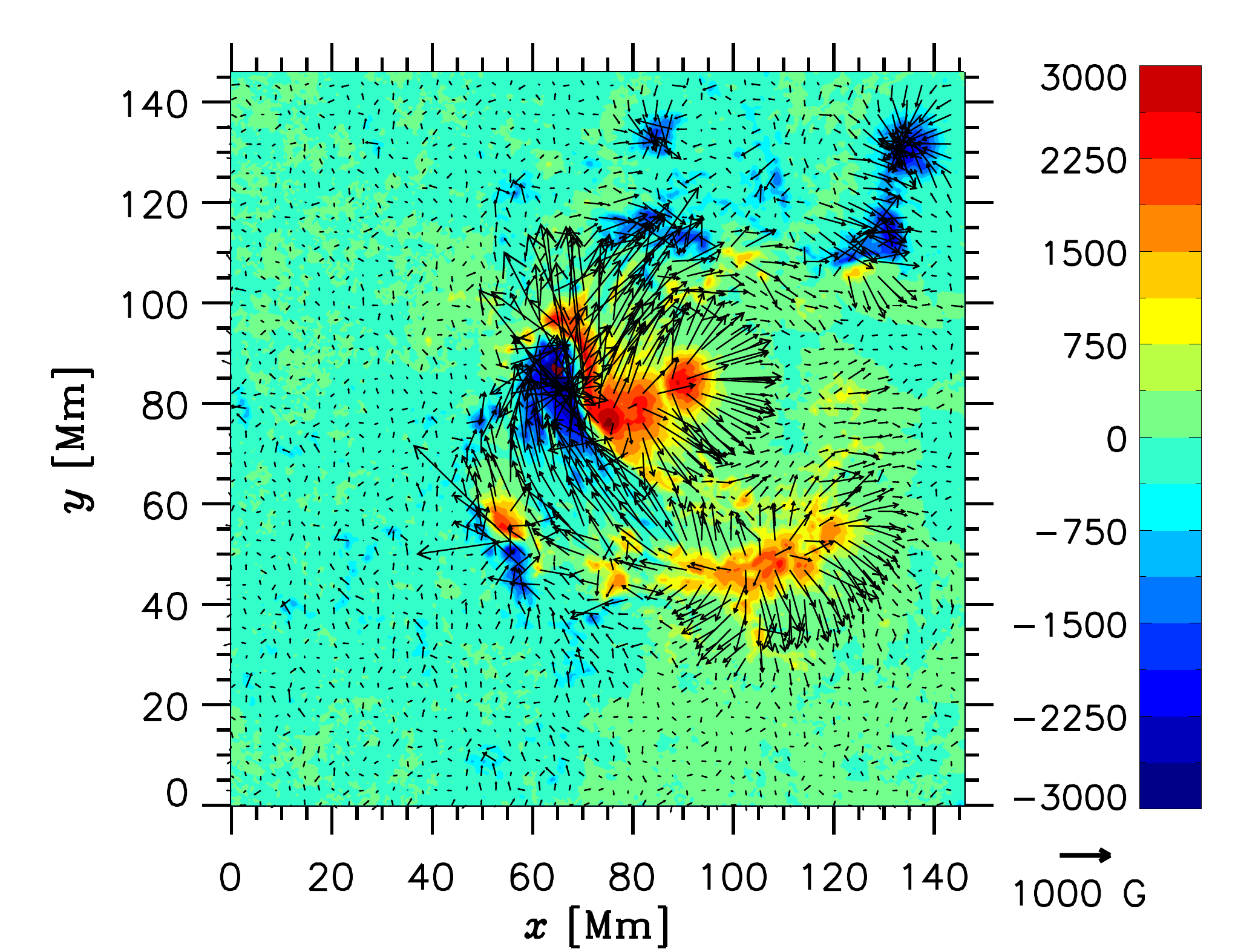}\\[-2pt]
\tiny{2017 September 06, 00:00 TAI}
\caption{Pattern of AR~12673 in the Carrington frame of reference at five times, not equally spaced (indicated under each row of panels). Left: optical images; right: magnetic field (colors representing the vertical component in gausses and arrows representing the horizontal component).}
\label{evolution}
\end{figure*}

\section{Observations and Data Processing}\label{obs}

Our study is based on data from the Helioseismic and Magnetic Imager (HMI) of the \emph{Solar Dynamics Observatory (SDO)}, which are stored at and available from the Joint Science Operations Center (JSOC, \url{http://jsoc.stanford.edu}). Specifically, we use a Spaceweather HMI Active Region Patch \citep[SHARP; see][]{Bobra_etal_2014} with the data remapped to a heliographic Lambert cylindrical equal-area projection (CEA). This automatically selected patch is centered at the flux-weighted centroid of the AR. The magnetic-field vector in the SHARP is decomposed into a radial (vertical), latitudinal and longitudinal components. The Dopplergrams are also CEA-remapped but not projected, still representing the line-of-sight, rather than radial, velocity component. We neglect the projection effects taking advantage of the fact that the AR was not far from the disk center on September 2--4, and the difference between the line-of-sight and the radial (vertical) component was not important.

As can be found from the headers of the downloaded FITS data files, the Carrington longitude of the center of this patch changed by 2.8 heliographic degrees over the 8-day period considered (2017 August 31, 00:00 TAI--2017 September 8, 00:00 TAI), the latitude being naturally constant. Alternatively to the SHARPs, JSOC offers a possibility of selecting, for a given AR, CEA-remapped arbitrarily sized continuum-intensity image patches with a fixed Carrington longitude; such images can be downloaded with time cadences down to 45~s---in contrast to the SHARPs, which have a cadence of 12~minutes. We use the short-cadence patches to construct full-vector velocity maps with the horizontal velocities computed by a modified local-correlation-tracking (LCT) technique \citep{gbuch}. To jointly analyze the two sets of images, we cut out fragments coaligned with the short-cadence patches from the SHARPs, using the values of the Carrington longitude indicated in the FITS headers.

The pixel size of the HMI images is 0.5 arcsec $\approx$ 366 km. The SHARP under study originally measures 547 $\times$ 372 pixels, or 200 $\times$ 136 Mm$^2$, and the size of short-cadence patches and our SHARP cutouts is 400 $\times$ 400 pixels, or 146.4 $\times$ 146.4 Mm$^2$.

We apply Fourier subsonic filtering \citep{Title_etal_1989} with a cutoff phase speed of 4 km\,s$^{-1}$ to the continuum images and Dopplergrams taken with a cadence of 45~s. To eliminate the velocity fluctuations on a granular scale, we smooth the line-of-sight velocities and reduce each smoothed Dopplergram to zero average.

The LCT procedure is applied here to a sequence of images selected with a cadence of 135~s. For this procedure to be successful, we magnify the images doubling the number of pixels in each horizontal dimension with the use of a standard bilinear-interpolation procedure. To obtain final representations of the horizontal-velocity field, we either average the measured velocities over nine time steps (20~minutes~15~s), which yields vector velocity maps, or integrate the displacements of imaginary corks, or test particles, distributed over the area of interest, thus constructing cork trajectories for 2-hour time intervals.

We will be interested in assessing the proper motions of the two components of the AR. However, the motion of cluster (2), consisting of a multitude of small features, can hardly be inferred from the LCT-based velocity field, which gives an idea of the local flow structure but does not yield the bulk velocity of the cluster. This is why, along with employing the LCT technique, we estimate horizontal photospheric velocities by tracking the motion of some clear-cut features in the images, such as pores, umbral islands, and penumbral fragments. To this end, we visually identify these in a sequence of images and read their coordinates. The advantage of this approach stems from the possibility of choosing most representative markers of plasma motion and obtaining results less affected by (convective) noises.

\section{Evolution of the active region}

Figure \ref{evolution} gives an overview of the evolution of AR~12673 as seen in white-light images and in magnetic-field vector maps (note that the both are not equally spaced in time). The coordinates $x$ and $y$ are measured in the longitudinal and the latitudinal direction, respectively. During the first two days after 2017 August~31, 00:00 TAI, the starting point of our consideration, the AR is represented in essence by only one isolated, well-developed positive-polarity sunspot (which has apparently survived two solar rotations that have elapsed after the decay of AR~12665); remember that we denote it as component (1) of the AR. It has a fairly regular structure with a nearly axisymmetric magnetic field, which exceeds 2300~G at the spot center (see the upper panel of Figure \ref{evolution}). In the Carrington frame of reference, the spot shifts eastward. It is remarkable in terms of its stability: it largely preserves its structure even at the later stage of strong interaction with cluster (2) consisting of smaller, less regular features.

\begin{table*}[!h]
\centering
\caption{Summary of the features tracked}\label{feat}
\begin{tabular}{cccccccc}
  \hline
  \hline
  Feature & $t_\mathrm{start}$ & $t_\mathrm{fin}$ & $x_\mathrm{start}$ & $x_\mathrm{fin}$ & $y_\mathrm{start}$ & $y_\mathrm{fin}$ & $\Delta x/\Delta t$\\
  no. &     (hr) & (hr)     & (Mm)   & (Mm)   & (Mm)   & (Mm) & (Mm\,hr$^{-1}$)   \\
  \hline
  \ \ 0 & \ \ \ \ 0  &    191 & 90.0 & \ \ 85.6 & \ \ 91.1 & \ \ 83.0 & $-0.023$\\
  \ \ 1 & \ \ 70     & \ \ 85 & 79.4 & \ \ 89.6 & \ \ 68.8 & \ \ 72.8 & 0.68\\
  \ \ 2 & \ \ 62     &    109 & 79.4 &    101.0 & \ \ 50.1 & \ \ 33.7 & 0.46\\
  \ \ 3 & \ \ 89     &    100 & 80.5 & \ \ 98.0 &    118.9 &    124.0 & 1.59\\
  \ \ 4 & \ \ 82     & \ \ 94 & 76.8 & \ \ 83.4 & \ \ 97.7 & \ \ 98.8 & 0.55\\
  \ \ 5 & \ \ 89     &    101 & 66.2 & \ \ 72.1 &    114.9 &    122.2 & 0.49\\
  \ \ 6 & \ \ 82     & \ \ 97 & 70.6 & \ \ 57.1 & \ \ 60.0 & \ \ 61.5 & $-0.90$\phantom{0}\\
  \ \ 7 & \ \ 84     &    118 & 53.4 & \ \ 48.7 & \ \ 84.5 & \ \ 79.0 & $-0.14$\phantom{0}\\
  \ \ 8 & \ \ 95     &    110 & 62.6 & \ \ 62.2 & \ \ 80.1 & \ \ 93.6 & $-0.027$\\
  \ \ 9 &    106     &    130 & 78.6 & \ \ 94.7 & \ \ 53.0 & \ \ 45.7 & 0.67\\
     10 &    109     &    132 & 79.4 & \ \ 89.6 & \ \ 98.4 &    109.4 & 0.44\\
     11 &    115     &    136 & 47.6 & \ \ 44.3 &    105.7 &    102.1 & $-0.16$\phantom{0}\\
     12 & \ \ 95     &    136 & 90.0 &    104.3 & \ \ 42.4 & \ \ 36.9 & 0.35\\
     13 & \ \ 94     &    106 & 80.8 & \ \ 75.0 &    107.9 &    104.3 & $-0.48$\phantom{0}\\
     14 &    107     &    119 & 72.1 & \ \ 69.1 & \ \ 77.5 & \ \ 81.2 & $-0.25$\phantom{0}\\
     15 & \ \ 84     & \ \ 97 & 70.6 & \ \ 83.8 &    114.5 &    125.5 & 1.02\\
  \multicolumn{7}{l}{Average of positive $x$-velocity values:} & $0.69\pm 0.27$\\ 
  \multicolumn{7}{l}{Average of negative $x$-velocity values (feature 0 excluded):} & $-0.33 \pm 0.24$\phantom{0}\\
  \multicolumn{7}{l}{Average of all $x$-velocity values (feature 0 excluded):} & $0.29 \pm 0.49$\\
  \hline
\end{tabular}
\end{table*}

\begin{figure*}[!h]
\centering
\includegraphics[width=0.35\textwidth,bb=30 10 420 540,clip]{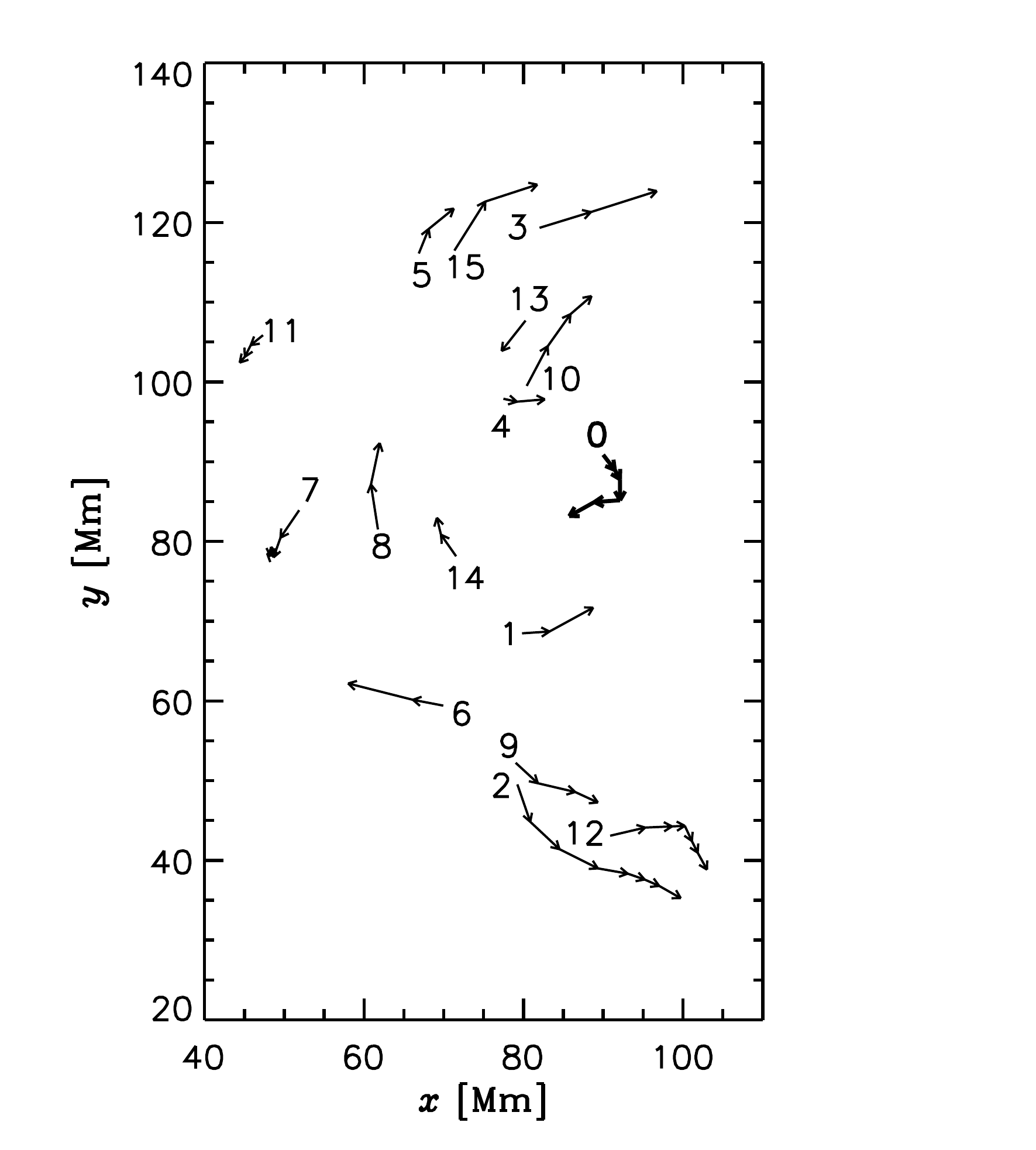}
\includegraphics[width=0.575\textwidth,bb=25 6 700 570,clip]{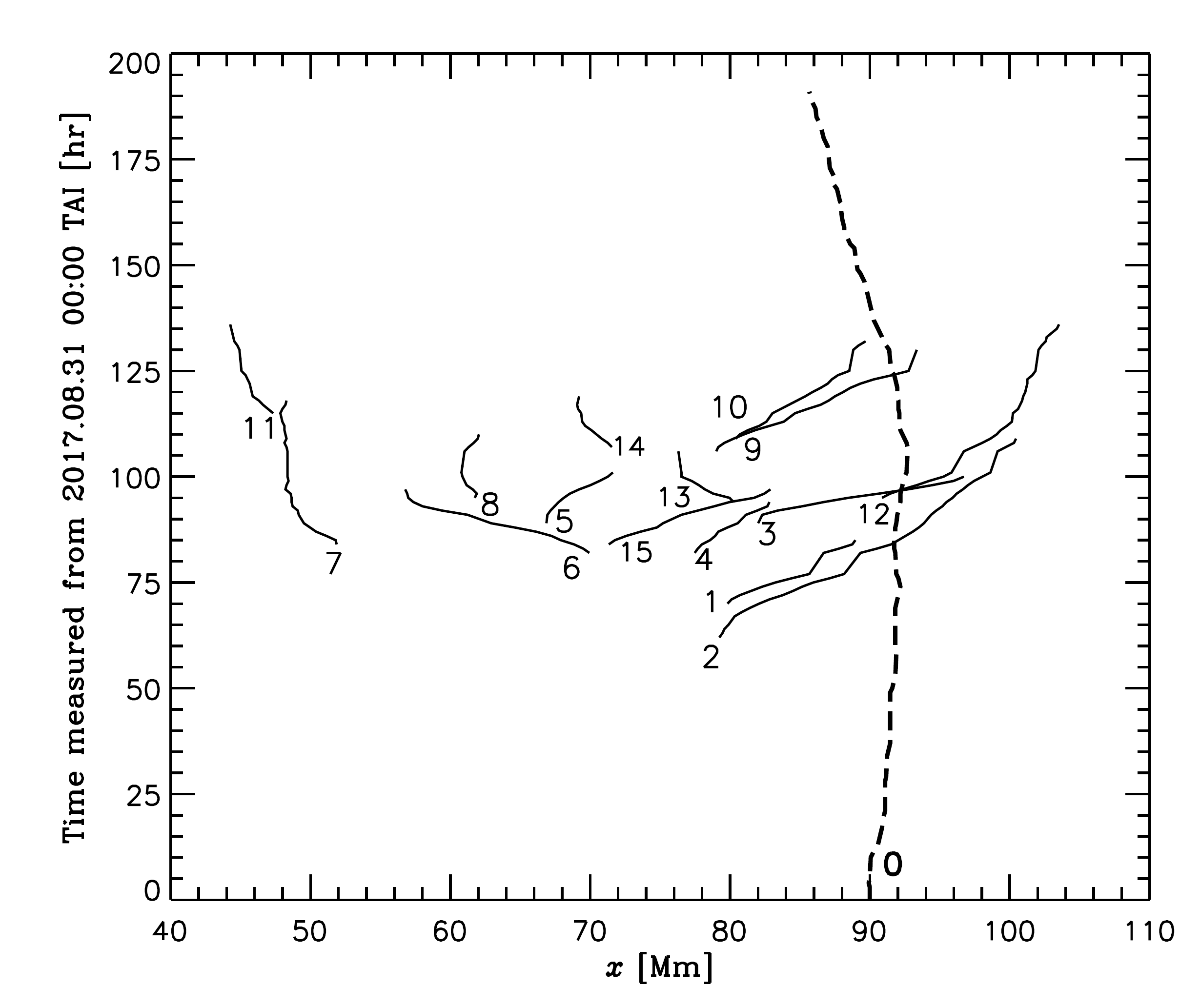}\\
\small{\hspace{-1.5cm}(a)\hspace{8.5cm}(b)}
\caption{(a) Map of drifting-feature tracks in the longitude--latitude plane; (b) longitude--time diagram of the drift of features.}
\label{graph}
\end{figure*}

On September~2, cluster (2) starts developing south--southeast of sunspot (1). A multipolar magnetic pattern emerges together with small spots, pores, and fragments of penumbra (panel for September~3, 00:00, in Figure~\ref{evolution}). During September~3, the cluster extends both longitudinally and especially latitudinally, acquiring a specific {arc-shaped} structure. Two elongated, bow-shaped chains of spot elements skirting around the major sunspot (1) originate, with two chains of magnetic elements spatially coinciding with them. The western chain is of the positive polarity, and the eastern one, of the negative polarity (see panel for September~3, 18:00). We will designate this evolutionary stage as the \emph{two-arc stage}  . These features persist over a certain time interval, the magnetic field being  enhanced but the whole pattern experiencing only moderate complications in its structure (see panel for September~4, 00:00). At later times, the interaction between AR components (1) and (2) becomes so strong that the pattern of sunspots and magnetic field undergoes dramatic changes (panel for September~6, 00:00). It is worth noting that the first flare in this AR occurred near the end of the two-arc stage (on September~3, at 20:45), while all other flares took place after a significant complication of the magnetic field, starting from September~4, 22:10.

It is the two-arc stage of the evolutionary scenario that will be of particular interest to us. Before and during this stage, the pattern of sunspots and magnetic field remains relatively ordered and can give some insight into the underlying physical mechanisms.

Even a visual inspection of the panels of Figure~\ref{evolution} for September~3, 18:00, and September~4, 00:00, reveals that AR components (1) and (2) are in relative motion, coming close to each other. To quantify this motion, let us keep track of some features identifiable with most certainty in the optical images of the AR. Information on these features is summarized in Table~\ref{feat}; for each feature, it presents the initial, $t_\mathrm{start}$, and the final, $t_\mathrm{fin}$, time of the interval for which the feature was tracked (in hours, measured from August~31, 00:00), the initial $(x_\mathrm{start},y_\mathrm{start})$ and the final $(x_\mathrm{fin},y_\mathrm{fin})$ local coordinates of the feature (in Mm) in the short-cadence patches (fixed to the Carrington frame of reference), and the mean $x$-velocity of the feature (in Mm\,hr$^{-1}$) defined as $\Delta x/\Delta t$, where $\Delta x = x_\mathrm{fin}-x_\mathrm{start}$ and $\Delta y = t_\mathrm{fin}-t_\mathrm{start}$. The movements of these features are also shown by Figure~\ref{graph}a in the form of tracks on the longitude--latitude plane in the Carrington frame of reference and by Figure~\ref{graph}b in a longitude--time diagram, with each track labeled by the feature number in both panels.

Feature 0 is the main sunspot, or component (1) of the AR. Its track is shown by a heavy curve both in Figure~\ref{graph}a and Figure~\ref{graph}b. This spot exhibits only a moderate shift in the Carrington frame of reference with a mean $x$-velocity of about $-0.023$~Mm\,hr$^{-1}$. The tracks of other features mainly refer to the two-arc stage and to the stage of subsequent complication of the entire pattern, September~3 and 4 (times of 72--120~hr). What attracts one's attention in Figure~\ref{graph}b is the fact that most features in the right-hand part of the diagram (region $x>65$~Mm) move westward, while a few features (6, 7, 11, 13, 14) in its left-hand part exhibit eastward drift. The entire flow pattern appears more consistent as considered in the longitude--latitude plane (Figure~\ref{graph}a). It becomes clear that these movements are predominantly directed along the arcs, roughly, to the northwest in the northern part of the pattern and to the southwest in its southern part (more precisely, the separatrix between the northward and southward streams is directed to the southeast of the main spot). Very few features exhibit arc-aligned movements directed oppositely compared to nearby features, so that countercurrent streams are intermixed to a certain extent: feature 13 moves oppositely compared to nearby features 3, 5, 10, 15, and features 7 and 11 move oppositely compared to feature 8. There are also motions across the arcs (features 1, 4, 6); they can be revealed more clearly from velocity maps and will be discussed below.

While feature 0 (the main sunspot) slightly shifts eastward, the cluster comprising other features moves as a whole westward, as can be grasped from Figure~\ref{graph}. This is apparent despite the presence of inner motions superposed on the bulk motion of the cluster. The data presented in Table~\ref{feat} and visualized in Figure~\ref{graph} make it possible to estimate the relative velocity of spot (1) and cluster (2). Comparatively stable and well localized features are few in number, and averaging their velocities implies fairly large mean absolute deviations (they are indicated in the table). Nevertheless, a figure of order 0.3~Mm\,hr$^{-1}$ seems to be a realistic estimate for the westward velocity of the cluster relative to the Carrington frame of reference and to the main spot.

\begin{figure}
\ \includegraphics[width=0.46\textwidth,bb=20 0 570 430,clip] 
{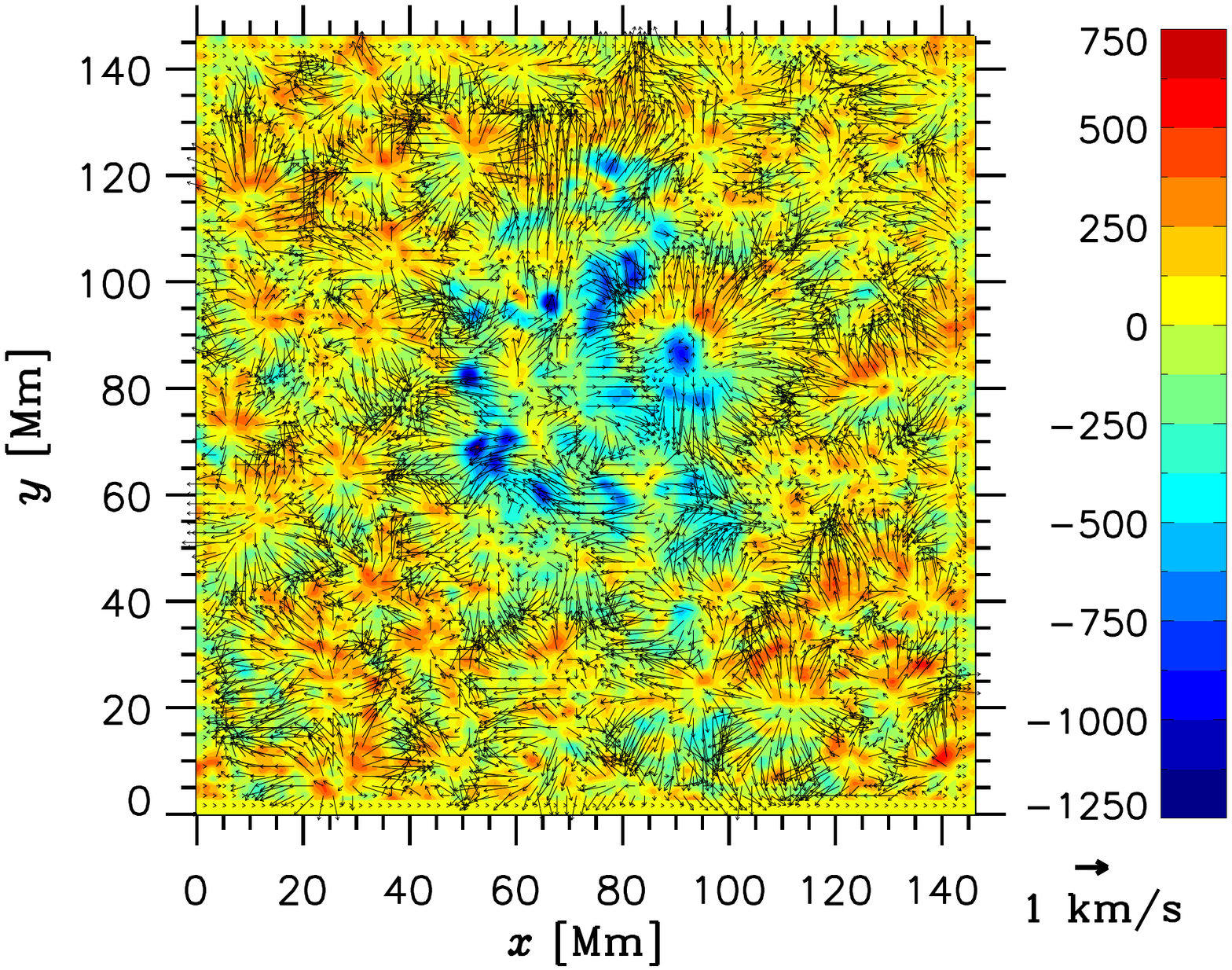}\\[-0.8cm]
\begin{center}{\small{(a)}}\\ \end{center}\vspace{-0.2cm}
\includegraphics[width=0.408\textwidth,bb=20 5 500 430,clip] 
{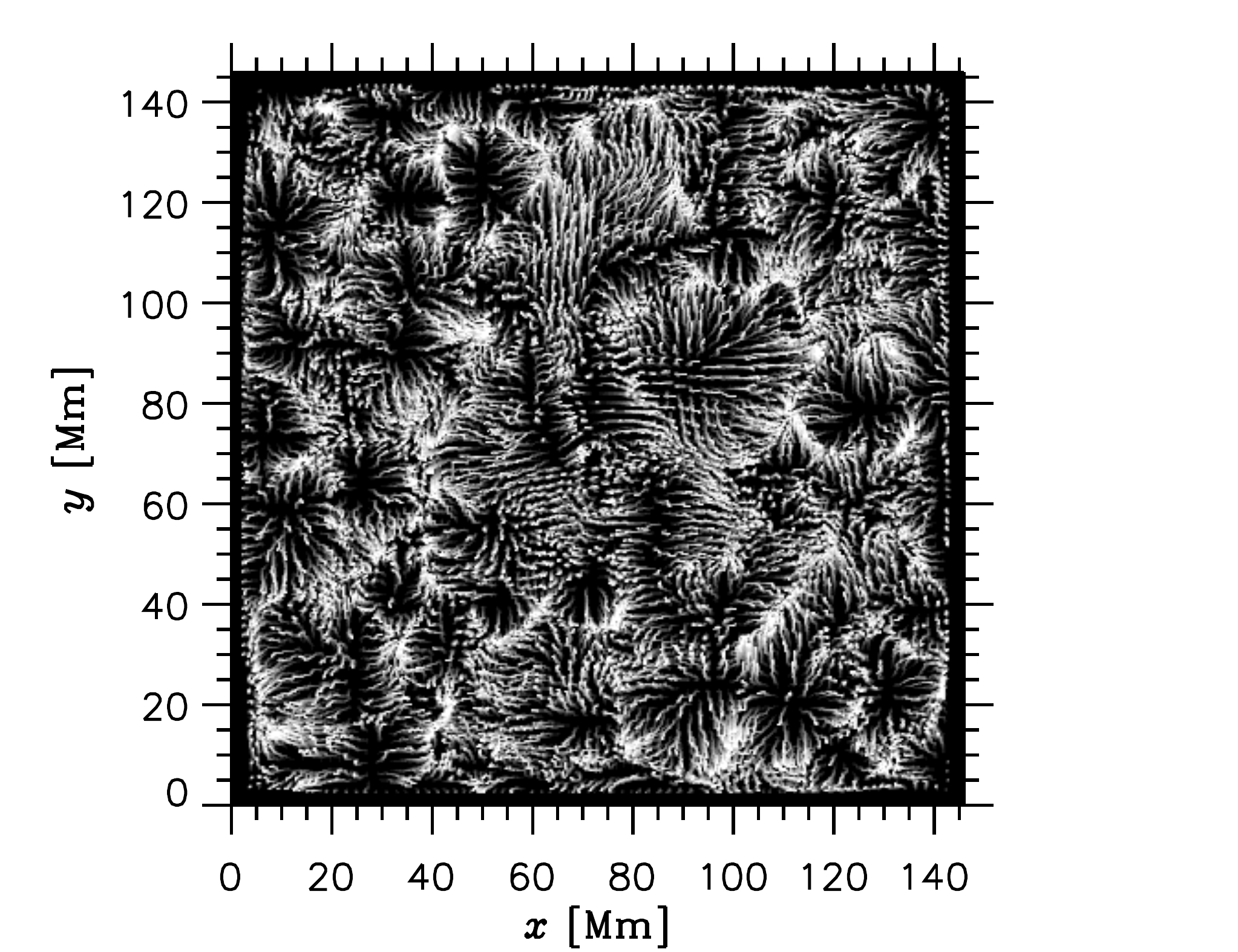}\\[-0.8cm]
\begin{center}\small{(b)}\\ \vspace{-0.05cm}
\includegraphics[width=0.24\textwidth,bb=0 0 388 500,clip]{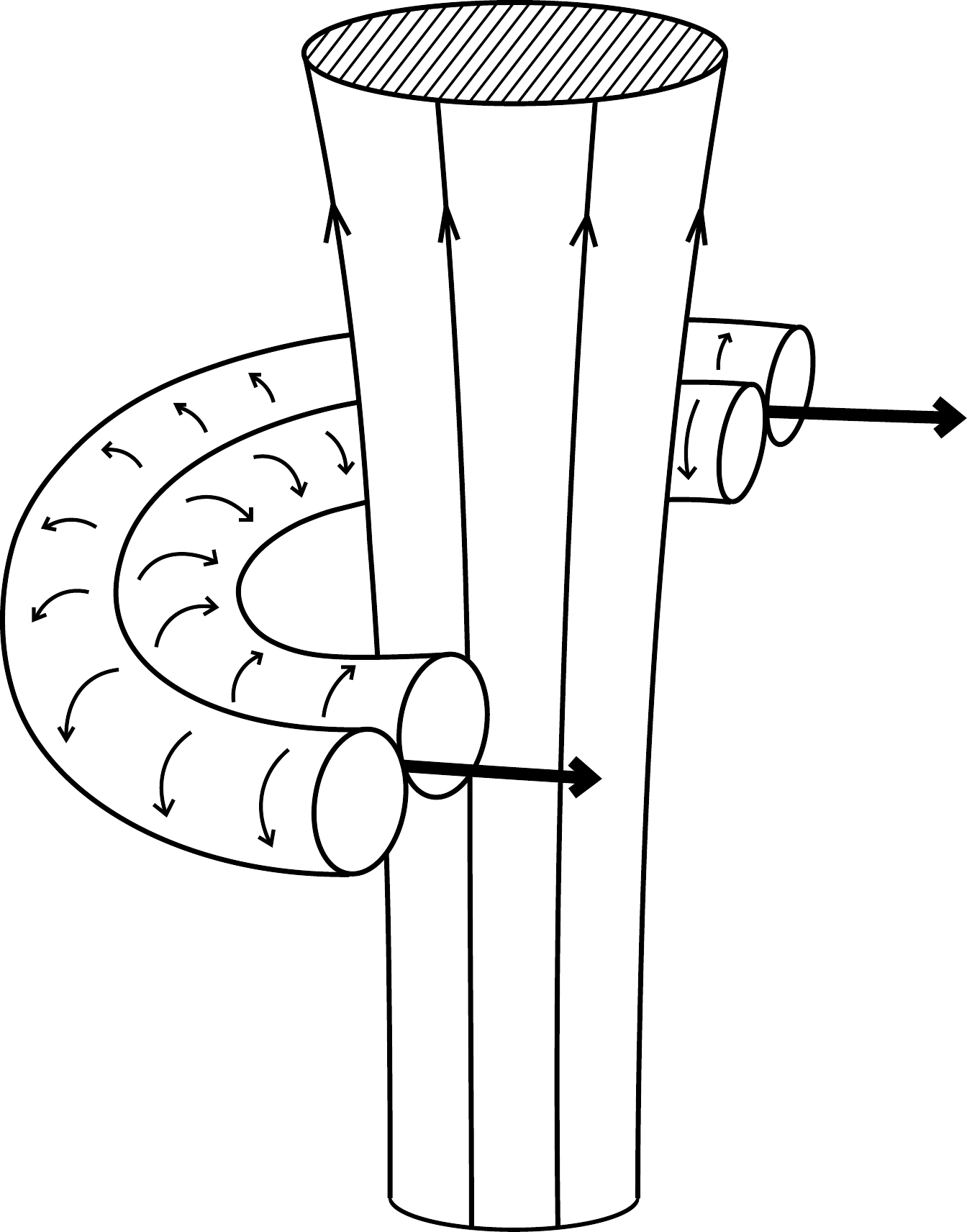}\\
\small{(c)}\end{center}\vspace{-0.5cm}
\caption{(a) Velocity field in AR~12673 at 2017 September 3, 13:22 TAI (colors representing the vertical velocity component in m\,s$^{-1}$ and arrows representing the horizontal velocity component averaged over an interval of 20 minutes 15~s centered at that time; the scale of the horizontal velocity is shown with a heavy horizontal arrow at the bottom right of the figure); (b) map of cork trajectories obtained by integrating the cork displacements over an interval of 2~hr 17 minutes centered near that time (the initial point of each trajectory is black and the final point is white, the brightness gradually increasing with time); (c) {schematic representation of the AR at the stage of the well-developed flow: the vertical column depicts the magnetic flux tube of the main sunspot, the roll-type flow is shown by short, curved arrows, and the velocity component along the rolls is displayed by heavy arrows.}}
\label{scheme}
\end{figure}

On the whole, as Figure~\ref{graph}a demonstrates, the pattern of motion of the features about the main spot bears amazing resemblance to the pattern of a \emph{fluid flow about a roundish body}. At the same time, the pattern of local inner motions in cluster (2), as obtained using the LCT technique, exhibits some likeness to the pattern of roll thermal convection. This is most pronounced at the two-arc stage of AR development and illustrated in Figure~\ref{scheme}a by a full-vector velocity map and in Figure~\ref{scheme}b by a map of cork trajectories: the flow pattern is like that of a couple of bowed convection rolls with superposed arc-aligned motions, as schematically sketched in Figure~\ref{scheme}c.

The indications for a roll flow are not very surprising by themselves: roll-like motions were previously detected in the solar photosphere \citep{Getling_2006,Getling_Buchnev_2008}. In our case, it is remarkable that (i) roll motions can be detected in the area where a relatively strong magnetic field with a fairly complex structure is present and (ii) the roll-flow pattern interacts with an obstacle moving relative to this pattern, i.e., with the coherent magnetic field of the main sunspot, and experiences deformations due to this interaction. Fine details complicating the pattern seemingly reflect a considerable role of the magnetic field affecting the flow.

\section{Physical Interpretation}

Our interpretation of the above-described evolution pattern of AR~12673 is based on the available information on the differential rotation of the Sun. We use data of helioseismological inversions determining the rotation rate of deeper layers. In particular, the difference in the rotation rate across the near-surface shear layer, or leptocline, can be estimated based on Figure~1 of \citet{Howe_etal_2000}. Specifically, at a latitude of $\varphi = \pm 9^\circ$, nearly corresponding to the latitude of AR~12673, the rotation frequency $\Omega/2\pi$ increases from the surface to depths of about $0.04R_\sun$ by $\Delta\Omega/2\pi\approx15$~nHz  (having characteristic values of about 460~nHz). The corresponding linear-velocity shear in a solidly rotating frame of reference is $\Delta u=\Delta\Omega\,R_\sun \cos \varphi = 0.24\ \mathrm{Mm\,hr}^{-1}$, which is a quantity of the same order of magnitude as the relative velocity of the main sunspot (1) and cluster (2), $0.3\ \mathrm{Mm\,hr}^{-1}$. This suggests that \emph{the main sunspot may be dynamically coupled to higher layers than the cluster}. The well-developed, stable magnetic feature---sunspot (1)---extends through the leptocline but moves with the surface layers, while the plasma of the deeper layers flows about this magnetic obstacle together with cluster (2). The long prehistory of sunspot (1) may be a fair indication that it has had time to become ``fixed'' to near-photospheric layers, whereas the fresh cluster magnetic field developing at deeper levels has not.

We can also put forward some tentative considerations of the magnetic-field emergence in cluster (2) during the two-arc stage. The observed pattern of this process resembles the interaction of a couple of convection rolls with a horizontal magnetic sheet lying at a certain depth. The field in this sheet should be directed from west to east. As it is carried by the upflow, which extends along the join between the two convection rolls, and then stretched by the diverging eastward and westward flows, is can form the negative and the positive arc observed along the downflow portions of the rolls.

\section{Note on the Formation of Light Bridges}

\begin{figure}
\centering
\includegraphics[width=0.342\textwidth, bb=26 0 440 400,clip] {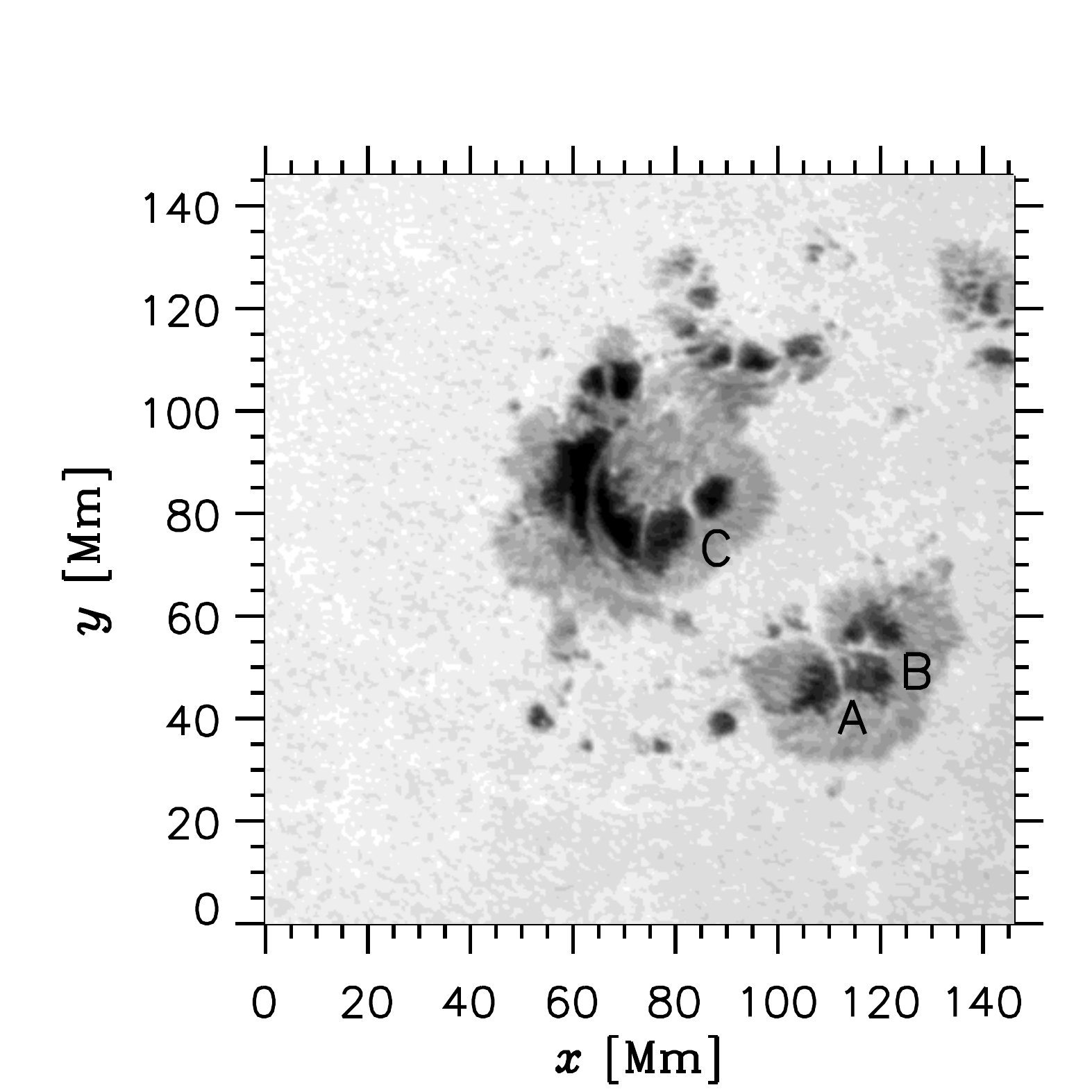}
\caption{Optical image of AR~12673 for 2017 September 07, 05:00 TAI, with three light bridges (labeled with letters) intersecting the umbrae of sunspots.}
\label{bridges}
\end{figure}

In addition, an interesting comment can be made {in passing} on the formation process of light bridges in sunspot umbrae. This phenomenon is typically attributed to the decay stage of sunspots \citep[see, e.g.,][where, in particular, a comprehensive survey of light-bridge properties is given]{Felipe_etal_2016}. AR~12673, which we consider here, demonstrates a different scenario of the development of light bridges. {Bridges A--C in Figure~\ref{bridges} formed as spot (1) and cluster (2) came close together to form a more complex sunspot; they appear as remnants of the bright space that previously separated the two components of the sunspot group. These bridges can be classified as \emph{strong} light bridges \citep{Felipe_etal_2016}: the magnetic field on both sides of them has the same polarity.}

\section{Summary and Conclusion}

We have analyzed the dynamics of AR~12673 using \emph{SDO}/HMI data for 2017 August 31, 00:00 TAI--2017~September 8, 00:00 TAI. The sunspot group in this AR consisted of (1) an old,  well-developed and highly stable, coherent sunspot and (2) a rapidly developing cluster of umbral and penumbral fragments. The most remarkable manifestation of the cluster evolution is the formation of two elongated, {arc-shaped} chains of spot elements skirting around the major sunspot (1), with two chains of magnetic elements spatially coinciding with the arcs (the two-arc evolutionary stage). The western chain is of the positive polarity, and the eastern one, of the negative polarity. Components (1) and (2) are in relative motion, coming close to each other; in its westward motion, cluster (2) is overtaking spot (1) at a speed of order 0.3~Mm\,hr$^{-1}$.

On the whole, the pattern of motion of the features about the main spot bears amazing resemblance to the pattern of a fluid flow about a roundish body. Together with the estimate of the relative speed of components (1) and (2) and in view of the existence of a near-surface shear layer, or leptocline, this suggests that the main spot may be dynamically coupled to higher layers than the cluster. According to helioseismological data, at the latitude of AR~12673, in a solidly rotating frame of reference, the linear-velocity shear  between the photosphere and the base of the leptocline (a depth of $0.04R_\sun$) is 0.24~Mm\,hr$^{-1}$. A velocity increase, relative to the photosphere, of the order of the estimated relative speed of sunspot (1) and cluster (2) should therefore be achieved at depths comparable to the leptocline thickness. Likely, the well-developed, stable magnetic feature---sunspot (1)---extends through the leptocline but moves with the surface layers, while the plasma of the deeper layers flows about this magnetic obstacle together with cluster (2). The long prehistory of spot (1) may be fair indication that it has had time to become ``fixed'' to near-photospheric layers, whereas the fresh cluster magnetic field developing at deeper levels has not. It seems reasonable to find out in further studies whether relatively low revolution rates is a distinctive property of old, stable sunspots.

The pattern of local inner motions in cluster (2) exhibits some likeness to the pattern of roll thermal convection. The observed pattern of magnetic-field emergence in cluster (2) during the two-arc evolutionary stage resembles the interaction of a couple of convection rolls with a horizontal magnetic sheet lying at a certain depth.

During the evolution of AR~12673, a few light bridges develop in the sunspot group. The formation of some of them is definitely not related to the decay of sunspots; in contrast, they are remnants of the bright space between the spots that have come close to each other. These bridges can be classified as strong light bridges: the magnetic field has the same polarity on both sides of them.

We see that the pattern of relative motion of different photospheric features may prove to be informative about the interaction of different near-photospheric layers. The diagnostic value of the flow patterns seems to deserve further investigation, in particular, with invoking helioseismological data.

\begin{acknowledgements}
The observational data were used here by courtesy of NASA/SDO and the HMI science teams. I am grateful to A.~A.~Buchnev, who developed the modified-LCT and the cork-tracking procedure, to K.~S.~Ganin for the preparation of Figure~\ref{scheme}c, and to B.~Yu.~Yushkov for a useful comment.
\end{acknowledgements}

\bibliography{Getling}

\end{document}